\documentclass[twocolumn]{aa}
\usepackage{graphicx}
\usepackage{txfonts}

\newcommand{\xmm}{{\sl XMM-Newton}}

\def\arcm{\hbox{$^\prime$}}
\def\arcs{\hbox{$^{\prime\hskip -0.1em\prime}$}}

\def\eg{{\it e.g.,~\/}}

\def\ie{{\it i.e.,~\/}}

\def\deg{{${ }^{\circ}$}}
\begin{document}
\title{The \xmm~slew survey in the 2--10 keV  band}

\subtitle{}

\author{R.S. Warwick
          \inst{1}
          \and
          R.D. Saxton\inst{2}
          \and
          A.M. Read\inst{1}
}


\institute{Dept. of Physics and Astronomy, University of Leicester, 
Leicester LE1 7RH, U.K.\\
\email{rsw@star.le.ac.uk}
         \and
XMM SOC, ESAC, Apartado 78, 28691 Villanueva de la Ca\~{n}ada, Madrid
              , Spain\\
}

\date{Received; accepted }

\abstract
{The on-going XMM-Newton Slew Survey (XSS) provides coverage of a significant
fraction of the sky in a broad X-ray bandpass. Although shallow
by contemporary standards, in the ``classical'' 2--10 keV band of X-ray 
astronomy, the XSS provides significantly better
sensitivity than any currently available all-sky survey.}
{We investigate the source content of the XSS, focussing on detections in
the hard 2--10 keV band down to a very low 
threshold ($\ge 4$ counts net of background). At the faint end, the 
survey reaches a flux sensitivity of roughly $3 \times 10^{-12} 
\rm~erg~cm^{-2}~s^{-1}$ (2-10 keV).} 
{Our starting point was a sample of 487 sources detected in the XSS (up to and including
release XMMSL1d2) at high galactic latitude in the hard band. Through 
cross-correlation with published source catalogues from surveys spanning the 
electromagnetic spectrum
from radio through to gamma-rays, we find that 45\% of the sources have likely
identifications with normal/active galaxies. A further 18\% 
are associated with other classes of X-ray object 
(nearby coronally active stars, accreting binaries, clusters of galaxies), 
leaving 37\% of the XSS sources with  no current identification.  
We go on to define an XSS extragalactic sample comprised of 219 galaxies and active
galaxies selected in the XSS hard band. We investigate the properties
of this extragalactic sample including its X-ray log N - log S distribution.}
{We find that in the low-count limit, the XSS is, as expected, strongly affected
by Eddington bias.  There is also a very strong bias in the XSS against the detection of 
extended sources, most notably clusters of galaxies. A significant fraction of the detections
at and around the low-count limit may be spurious.  
Nevertheless, it is possible to use the XSS to extract a reasonably robust 
sample of extragalactic sources, excluding galaxy clusters. The differential 
log N - log S relation of these extragalactic
sources matches very well to the HEAO-1 A2 all-sky survey measurements  at 
bright fluxes and to the 2XMM source counts at the faint end.}
{The substantial sky coverage afforded by the XSS makes this survey a 
valuable resource for studying X-ray bright source samples, including those selected
specifically in the hard 2--10 keV band.}

\maketitle


\section{Introduction}

One of the on-going programmes of {\it XMM-Newton} is a shallow large-area 
X-ray survey based on data recorded by the on-board cameras as the observatory 
slews from one target source to the next. The first catalogue of {\it XMM-Newton} 
Slew Survey sources (XMMSL1) was released in May 2006 and 
then updated in August 2007 (XMMSL1d1).  Later additions
to the catalogue were made in  April 2008 (XMMSL1d2), July 2009 (XMMSL1d3), 
April 2010 (XMMSL1d4) and most recently June 2011 (XMMSL1d5).  The full 
{\it XMM-Newton} slew catalogue
is available from the {\it XMM-Newton} Science Archive (XSA). 

As discussed by Saxton et al. (2008), the {\it XMM-Newton} Slew Survey (hereafter XSS)
is based solely on data from the EPIC pn camera, since the longer readout times of
the MOS cameras lead to a very elongated point spread function in slew observations.
The in-orbit slew speed of 90 degrees per hour results in an exposure time for sources
lying on the slew path of between 1--11s.  The XSS records the count rates of sources
in two nominal energy ranges, namely a ``soft'' 0.2--2 keV band and 
a  ``hard'' 2--12 keV band.  
In practice, the high energy fall-off in the pn detector response results in
a negligible contribution to the XSS source count rates from X-ray photons
with energies between 10--12 keV and, hereafter, we take the energy range
of the XSS hard band to be 2--10 keV.

In the soft 0.2--2 keV band  the limiting sensitivity for 
source detection in the XSS is roughly $6 \times 10^{-13} \rm~ergs~cm^{-2}~s^{-1}$,
which is very comparable to that reached in the ROSAT All-Sky Survey (RASS; \cite{voges99}).
In the hard 2--10 keV band the limiting sensitivity for source 
detection is approximately $3 \times 10^{-12} \rm~ergs~cm^{-2}~s^{-1}$. This is roughly an
order of magnitude deeper than the {\it all-sky} surveys 
which are currently available in this energy range, such as those from 
{\it Uhuru}, {\it Ariel V}, and {\it HEAO-1} 
(\cite{forman78}; \cite{mchardy81}; \cite{piccinotti82}; \cite{wood84}).
The situation is, of course, very different if one considers surveys 
with sky-coverage much less than $4\pi$ steradians. This includes the low/medium depth 
surveys from ASCA (\cite{ueda05}) and BeppoSax (\cite{giommi00}), 
the multi-observation 
dedicated surveys and serendipitous observations carried out by {\it XMM-Newton} 
and {\it Chandra} (\eg \cite{mateos08}; \cite{elvis09}) and the ultra-deep pencil-beam 
surveys also completed by these latter two missions (\eg \cite{alexander03};
\cite{luo08}; \cite{xue11}; \cite{brunner08}; \cite{comastri11}). These recent surveys 
reach sensitivity levels three orders of magnitude fainter than  the  XSS.

Within this setting, the XSS survey has two important attributes: (i) the coverage
of a relatively large sky area (35\% of the sky up to and including the
XMMSL1d2 release considered here, rising to 52.5\%  with the most recent
increments) and (ii) the sampling of a flux range in the hard band for
which current source catalogues provide far from comprehensive information. At the
high flux end, the XSS overlaps with the existing all-sky 2--10 keV surveys, whereas
at its sensitivity limit, the XSS is detecting objects which are still ``X-ray bright'' 
in the context of the targeted \xmm~programme.   Although well represented in the 2XMM 
catalogue (\cite{watson09}), the use of this catalogue (or similar catalogues) to evaluate
the statistical properties of ``bright'' sources is greatly complicated by the complex
biasses and selection effects, which necessarily influence the target and field selection 
of an observatory-class mission (\cite{mateos08}).

In this paper we explore the statistical properties of the XSS, focussing on sources
detected in the hard band. More specifically we investigate the extragalactic
component of the XSS  catalogue. In the next section, we describe the criteria
used to select our source sample, examine some of the basic properties
of the XSS sources, investigate the likely false detection rate and also
explore the strong bias against the detection of extended sources.
In \S 3 we describe the results of a cross-correlation of the XSS source
positions with published source catalogues spanning a wide range of wavebands
and where possible categorize each source on the basis of the type of
counterpart. We then go on in \S 4, to explore the multiwavelength properties
of an XSS hard-band selected sample of extragalactic sources
(excluding galaxy clusters). In \S 5 we construct the log N - Log S
distribution of the extragalactic sample utilizing the results 
from a simulation of the survey process. We then discuss the degree to which
our XSS source sample bridges the gap in flux coverage between the 
``classical'' 2--10 keV all-sky surveys and the current generation of
deep pointed-mode surveys. Finally in \S6 we provide a brief summary of
our conclusions.

\section{The XSS hard-band selected sample}

\subsection{Definition of the source sample} 

Full details of the procedures used to construct a source catalogue based
on {\it XMM-Newton} slew data are reported in Saxton et al. (2008). Our 
starting point was the ``clean'' version of the XSS catalogue based on slews
performed up to Jan. 2008 (XMMSL1d2), which contains a total 
of 7686 sources (compared to the 2692 ``clean'' sources reported by 
Saxton et al. based on a more limited set of slew observations).  

When we select only those sources detected in the XSS hard band\footnote{For a 
source to be included in our sample we require a detection maximum likelihood 
($maxl$) in the XSS hard band of either $maxl > 10$, if the background 
count rate was less than 3 $\rm ct/s$, or $maxl > 14$ otherwise.
These are the same criteria as used to define the ``clean'' XSS sample,
except now applied specifically to the hard band. Note that in the on-line
XSS catalogue the {\it maxl} parameter has the designation DET\_ML.}, there is 
a dramatic drop to 796 catalogue entries.  
A further requirement that sources be located well off the 
Galactic Plane, with $|\rm b| > 10$\deg, results in a ``preliminary'' list 
of 617 hard-band selected sources.  The total area of 
high-latitude sky encompassed by the slew observations was 14233
square degrees, but after allowing for overlaps this figure reduced to
11914 square degrees (\ie 35\% coverage of the sky region above
$|\rm b| > 10$\deg).

The next step was to filter the source sample on the basis of various parameters
recorded in the XSS database. Selection criteria were set as follows: (i) a 
minimum net source count of 4 (\ie the number of counts assigned to the source 
after background subtraction); (ii) a minimum  hard-band count rate of 0.4 ct/s;  
(iii) a minimum effective exposure time of 1 s and (iv) a maximum source extent 
of 15 pixels ($\approx$ 60\arcs).  Application of these criteria reduced 
the sample to 540 sources.  We further applied a cut on the maximum count
rate of 10 ct/s; this serves as a precaution against pile-up effects in the detector 
but also proved a useful upper bound in the investigation of the source number counts. 
This removed detections of 4 well-known
active galactic nuclei (AGN), namely Mrk421,  
IC4329A, NGC5506 and 3C390.3, 16 bright Galactic X-ray binary sources
and one likely spurious source.
The result was a sample of 519 sources detected in the pn camera 
with count rates in the range 0.4-10 ct/s. The next consideration was 
duplicate detections of the same source.  There were
30 duplicates and one example of a triple detection in our list. 
We excluded duplicates by selecting the source detection with the
highest maximum likelihood in the XSS hard band\footnote {Note that a 
duplicate detection of Mrk 421 at a count rate less than the 10 ct/s upper threshold,
restored this source to our final sample.}. With the duplicates removed, our final sample
consisted of 487 hard-band selected sources.

\begin{figure*}
\begin{center}
\begin{tabular}{c}
\rotatebox{-90}{\includegraphics[height=14.5cm]{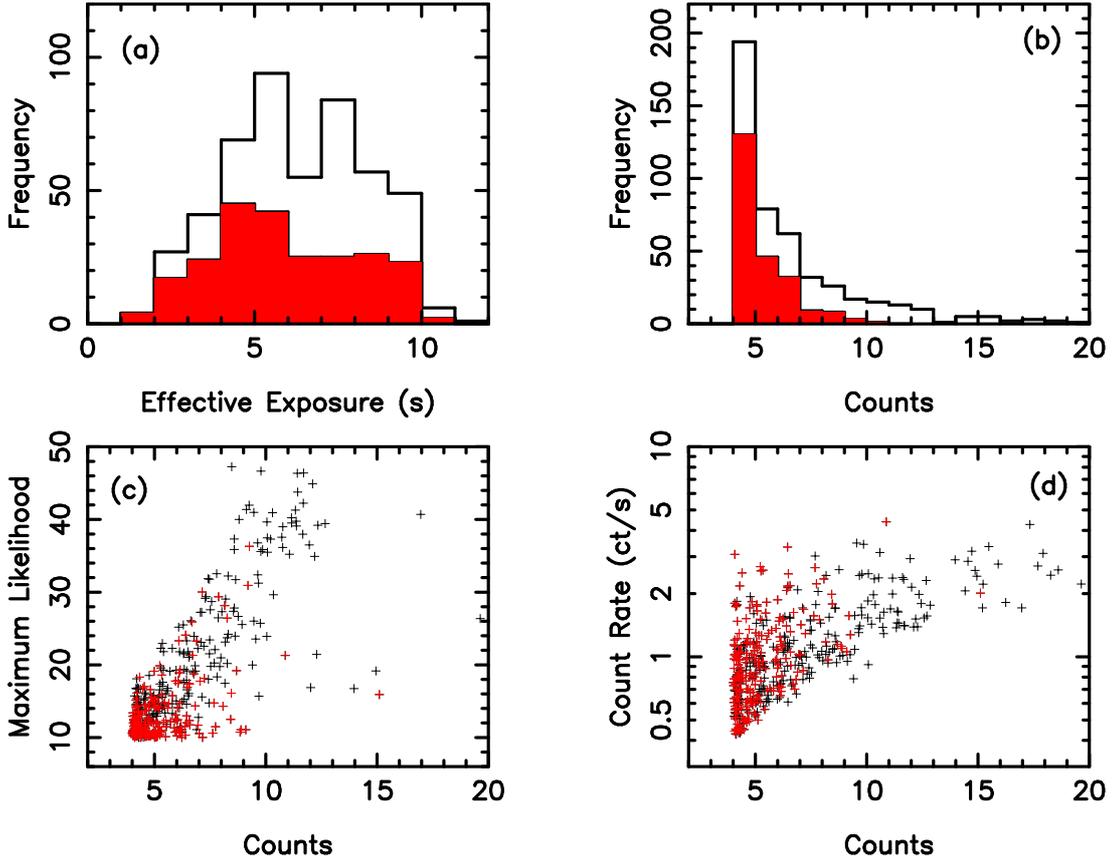}}
\end{tabular}
\end{center}
\caption
{Characteristics of the 487 sources comprising the XSS hard-band selected sample. 
{\it Panel (a):} Distribution of the effective exposure time of the full sample.
The filled (red) histogram corresponds to the subset of XSS sources which
were detected only in the XSS hard band. 
{\it Panel (b):} Distribution of the net counts recorded in the 
hard (2--10 keV) band (after background subtraction). The filled
(red) histogram again corresponds to the hard-band only sources.
{\it Panel (c):} Net counts versus the maximum likelihood of the detection
with the hard-band only sources plotted in red.
{\it Panel (d):} Net counts versus the corresponding count rate in $\rm ct/s$
with the hard-band only sources plotted in red.
For clarity the 20 sources with net hard counts in the range 20-70
are not shown in {\it Panels (b)-(d)}.
}
\label{fig:1} 
\end{figure*}

Some of the properties of this XSS source sample are illustrated in Fig. \ref{fig:1},
where a distinction is made between the full sample and the subset of the XSS 
sources which were detected solely in the hard band (see \S \ref{soft_det}
for further consideration of this issue). 
The bulk of the sources have exposure times in the range 2--10s (Fig. \ref{fig:1}a).
The distribution  of the counts recorded in the hard band (after
background subtraction) is strongly weighted towards the 4-count limit with 
194 sources having between 4--5 net counts (Fig. \ref{fig:1}b).
Fig. \ref{fig:1}c shows that the net counts and the maximum likelihood for the 
detection are, as expected, well correlated, albeit with a significant scatter.  
Finally Fig. \ref{fig:1}d shows the relation between the net counts and
the derived count rate; here the impact of the factor of 5  spread in the effective
exposure time is apparent in terms of the commensurate spread in the count rate at
a given net count.

If we  assume a source spectral model comprising
a power-law continuum with photon index $\Gamma = 1.7$ with soft X-ray absorption 
equivalent to a column density $N_H = 3 \times 10^{20} \rm~cm^{-2}$, then the count rate 
to flux conversion is  1 $\rm ct/s$ = $8.1 \times 10^{-12} \rm~erg~s^{-1}~cm^{-2}$ 
(2--10 keV) for the medium filter.  Our XSS hard-band selected sample therefore
includes sources with 2--10 keV fluxes broadly in the range
0.3--8 $\times 10^{-11} \rm~erg~s^{-1}~cm^{-2}$. 

\subsection{Concurrent hard- and soft-band detection}
\label{soft_det} 

\begin{figure}
\begin{center}
\begin{tabular}{c}
\rotatebox{-90}{\includegraphics[height=8.0cm]{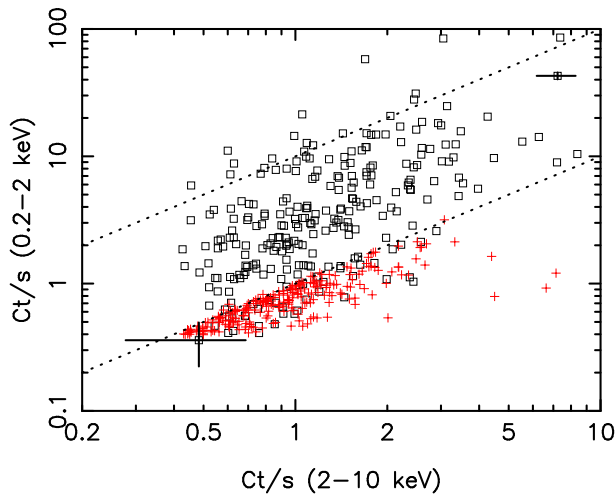}}
\end{tabular}
\end{center}
\caption
{XSS soft-band (0.2-2 keV)  count rate versus the hard band (2-10 keV)
count rate.  Sources detected in both the hard and soft bands are plotted
as squares, whereas those for which the soft-band measurement is an upper
limit (corresponding to 4 net counts) are shown as (red) pluses.  Error
bars are shown for one high and one low-count rate source for illustration. 
The dotted diagonal lines correspond to soft to hard count-rate ratios
of 1:1 and 10:1. 
}
\label{fig:2} 
\end{figure}

In addition to the 2--10 keV measurements, the XSS also provides
soft band (0.2--2 keV) data. Just over half (254 sources out of 487)
of the hard-band selected sources were also detected simultaneously in the
XSS soft band. Fig. \ref{fig:2} shows a comparison
of the count rates measured in the two XSS bands for the sources in
our sample. In this figure the lower and upper diagonal lines represent 
soft:hard  count-rate ratios of 1:1 and 10:1 respectively.
For a source spectral model comprising a power-law continuum with
$\Gamma = 1.7$  subject to soft X-ray absorption equivalent to $N_H = 3 \times 10^{20} 
\rm~cm^{-2}$ (the fiducial spectral form on which the count rate to flux calibration
is based), the soft to hard count-rate ratio is $\approx 2.7$.  This ratio 
increases to over 10 for~$\Gamma \approx 2.5$;  similarly it drops to below 1
if the column density is increased to $N_H \approx 5 \times 10^{21} \rm~cm^{-2}$
(for $\Gamma = 1.7$).

\subsection{False detections in the XSS}
\label{false}

A survey which extends to sources with only 4 net counts, measured against a low 
but not entirely negligible background, will include some false detections arising
from Poissonian noise spikes.  In the case of the XSS sample one would expect the
vast majority of the false detections to reside amongst the sources detected only
in the hard band (since the probability of a simultaneous false detection in
both the hard and soft bands
will be very small).  It is certainly true that the subset of ``hard-band only'' 
sources has a somewhat stronger weighting towards low net counts than the sources
detected in both bands (see Fig. \ref{fig:1}b). Specifically, 130 out of 233 (56\%)
of the hard-band only sources have between 4--5 net hard-band counts compared
to 64 out of 254 (25\%) for the dual-band detections.

To obtain an estimate of what fraction of the 233 XSS sources detected only in the
hard-band might actually be spurious detections, we have applied the following
argument. If we crudely approximate the source detection process (ignoring any
issues relating to the background subtraction) as a search for groupings 
of $N$ or more counts within a source cell of radius $R$ $=$ 20\arcs, 
then there are approximately $1.4 \times 10^{8}$ such cells
over the full XSS survey region. The typical background count in such a cell,
based on the actual backgrounds measured in the XSS images, was in
the range 0.01-0.05 counts, for which the Poissonian 
probability of measuring $N \geq 4$ is roughly $1.0 \times 10^{-7}$
(applying a weighted average over the background distribution). This implies
the detection of just 14 spurious sources. The predicted false detection rate
increases rapidly with $R$, with $R$ = 30\arcs~giving roughly 150 false
detections (although this is probably an over-estimate of the
effective source cell size). A further observation is that the low-count
sources are not preferentially drawn from slew datasets with higher
than average background; this suggests that if there is a high
false detection rate amongst such sources, then then its origin is
not solely due to Poissonian background fluctuations. 

As a further test we randomised the photon distribution in a subset of the
slew survey images and reran the source search algorithm. After applying 
the same selection criteria as employed in the current study, the number
of spurious source detections, scaled to the full XSS area, was 33. However,
we note that this approach does not encompass all of the complexities of the
source detection within the XSS.

We will return to this issue when we consider the high number
of ``unidentified'' sources amongst the subset of sources detected only
in the XSS hard band. 

\subsection{Bias against the detection of extended sources}

Some of the difficulties relating to the detection of extended sources in
the low-exposure, low-count regime of the XSS have been 
discussed by Saxton et al. (2008).  In practice, these issues  translate
to a fairly strong bias against the inclusion of extended objects
within the current XSS sample. This selection bias will particularly
impact on X-ray sources associated with nearby clusters of galaxies,
which form a substantial fraction of the high-latitude source population seen in 
{\it all-sky} surveys such as {\it Uhuru}, {\it Ariel V}, and {\it HEAO-1}.

In selecting our XSS sample we imposed an upper limit on the 
extent parameter of 15 pixels ($\approx$ 60\arcs). 
However, when we relax this selection criterion,
the total number of sources in the sample only increases by six, none of
which are obvious cluster candidates.  Given this outcome, it is
clear that the bias against extended objects is an intrinsic feature
of the XSS.

To illustrate the magnitude of the problem, we have employed 
the complete sample of high-latitude 
($|b| > 20$\deg) sources detected by the {\it HEAO 1} A-2 experiment
in the 2--10 keV band (\cite{piccinotti82}) as a comparator. 
This sample (which we hereafter refer to as the Piccinotti sample)
comprises 68 sources, 30 of which are identified as clusters of galaxies.
We have investigated whether the XSS encompasses the
positions of each of the Piccinotti sources and, where there
is coverage,  whether a hard-band detection resulted.  
For the non-detections we have gone back to the slew-survey data
and determined an upper limit using a Bayesian-based
on-line tool (\cite{saxton11}).

The results of this investigation were as follows. Ten 
of the Piccinotti clusters were covered by the XSS
(up to and including XMMSL1d2). Of these 3 appear
in our current sample with extent parameters in the
range 3-14 pixels. A further {\it HEAO 1} A-2 cluster was detected
with an extent parameter of 20 pixels (but in the event this source did
not make it into the XSS ``clean'' sample which was the starting point
of our selection process). The X-ray fluxes derived for these
4 objects from the XSS were comparable (within a factor $\sim 2$)
with those measured by the {\it HEAO 1} A-2 experiment.  
The remaining 6 clusters were not detected with upper-limits 
(assuming point-like sources) between 3-30 times lower than
those predicted from the {\it HEAO 1} A-2 fluxes. As a further comparison
the same analysis carried out for the 12 AGN in the Piccinotti
sample covered by the XSS gave
10 detections plus 2 upper-limits with inferred
variability factors typically in the range 1-3.

We conclude from this analysis that a strong bias does exist
against the detection of extended objects in the XSS.  

\section{Counterparts to the XSS sources}

\subsection{Identification process}

The next step was to draw together the available multiwaveband information relating 
to the likely counterparts of the sources in our XSS hard-band selected sample.
The starting point was the cross-correlation of the XSS positions with other source
catalogues encompassing a wide wavelength range.  Saxton et al. (2008) quote the
typical astrometric uncertainty of XSS positions to be 8\arcs~(the 68\% confidence
error radius), but for the purpose of cross-correlation with other catalogues,
we initially adopted a search radius of 30\arcs. This process relied heavily on the
facilities of
Vizier\footnote{http://vizier.u-strasbg.fr/viz-bin/VizieR/},
supplemented by reference to Simbad\footnote{http://simbad.u-strasbg.fr/simbad/}
and NED\footnote{http://ned.ipac.caltech.edu/}.
A few catalogues not currently available at Vizier
were also accessed directly from the survey websites, the most notable
being the Sloan Digital Sky Survey SDSS\footnote{http://www.sdss.org/} and
the Galaxy Evolution Explorer, GALEX\footnote{http://galex.stsci.edu/galexview/}, 
source catalogues. In a preliminary pass we made a judgement
on the likely identification of the X-ray source, if any.
In the great majority of cases this involved consideration of the available
information for a single putative counterpart, although in a small number of
instances it was necessary to select the most likely counterpart amongst two
or three candidates (using criteria such as the offset from the X-ray position
and the brightness of the sources). 

This first iteration separated the XSS sample into six broad categories. 
Sources of likely  extragalactic origin were flagged as either {\it AGN},
{\it Galaxies} or {\it Clusters} (of galaxies). In general sources
were categorized as {\it Galaxies} rather than {\it AGN} when there was a
report of an extended
optical/IR morphology and/or an associated redshift, but no definitive
confirmation of the presence of an active nucleus.
Similarly objects of likely Galactic origin were divided into {\it Stars} 
(mostly nearby objects with active stellar coronae)
and {\it Other} categories (including X-ray binaries, cataclysmic variables
supernova remnants and one pulsar). X-ray sources most likely residing
in Local Group galaxies (M31, LMC and SMC) were also assigned to the {\it Other}
category, as was an Ultraluminous X-ray Source (ULX) detected NGC 2403.  
Finally the XSS sources without an obvious counterpart were categorized
as {\it Unidentified}.

Subsequent iterations allowed the refinement of the above process
using the knowledge derived from the earlier cycle. In particular,
the great majority of the sources categorized as either the {\it AGN}, {\it Galaxies}
or {\it Stars} were found to have both near-IR and mid-IR counterparts in
the 2MASS\footnote{The Two Micron All Sky Survey (2MASS)  
(\cite{cutri03}; \cite{skrutskie06})
provides uniform coverage of the entire sky in three near-infrared 
(NIR) bands, namely in J (1.25$\mu$), H (1.65$\mu$) and K$_{s}$ 
(2.17$\mu$).}
and WISE\footnote{The Wide Field Infrared Explorer (WISE) has recently observed
the entire sky at 3.4, 4.6, 12 and 22 $\mu m$. The All-SKY Data Release covers
$>99\%$ of the sky and incorporates the best available calibrations
and data reduction algorithms (\cite{cutri12}).} surveys respectively.  
This allowed both the
astrometric precision of the XSS and the IR colours of the counterparts
to be explored in a systematic way.
One application of the latter was to resolve some initial ambiguity as 
to whether the likely counterpart was a galaxy or a star on the basis of
its IR colour (see \S \ref{multiwavelength}).
Also by exploiting the GALEX survey (\cite{martin05}), which currently
provides UV coverage of roughly two-thirds of the sky in the near-UV (NUV, 1350-1750 \AA)
and far-UV (FUV, 1750-2750 \AA) bands, a number of sources
initially categorized as {\it Galaxies} were switched to {\it AGN} on the basis
of their UV to near-IR colour (\S \ref{sample}).

An important consideration emerged in relation to the subset of XSS sources which were
detected only in the hard X-ray band. Hereafter we refer to this subset as the 
``hard-only'' sources, as opposed to the ``hard+soft'' sources detected simultaneously
in both XSS bands. For the hard-only sources the success rate
in finding relatively bright, plausible counterparts within a nominal search radius
proved to be surprisingly low.  As a consequence, in searching for putative 
counterparts down to fainter magnitudes, the rate of chance coincidences
emerged as a crucial issue. In the event it proved necessary to focus on the
hard-only sources with reasonable position errors as
defined by the XSS {\it pos\_err}~\footnote{The
{\it pos\_err} parameter provides an estimate of the uncertainty
of the XSS position, albeit a rather crude one for sources at the
low-count threshold. In the present work we restrict its use to
filtering out the hard-only sources with particularly
poor position determinations. Note that in the on-line XSS catalogue
the {\it pos\_err} parameter has the designation RADEC\_ERR.}. As a  practical approach
the 68 hard-only sources with {\it pos\_err} $> 10$\arcsec~were 
excluded from the identification statistics, although in practice
there were very few plausible counterparts amongst this group.
As a further step rather tight constraints were placed on the
positional offsets and limiting magnitudes of the objects
considered to be plausible counterparts to the remaining hard-only
sources (see \S\ref{hardonly} for further details).

A summary of the final identification statistics for the full XSS sample is
provided in Table \ref {table:1}, where the division is into the six broad
source categories defined earlier.
Table \ref {table:1} also provides the identification information 
split down to the hard+soft and hard-only source subsets.

\begin{table}
\caption{Division of the XSS sample into source types.}
\label{table:1}
\begin{center}
\begin{tabular}{l c c c}   
\hline
Source Type  & All   &  Hard+Soft & Hard-only  \\
\hline
AGN          & 181  &  160  & 21  \\
Galaxies       &  38  &   16  & 22  \\
Clusters      &  10  &   10  &  0  \\ 
Stars         &  53  &   51  &  2  \\
Other        &  27  &   17  & 10  \\
Unidentified & 178  &    0  & 178$^{\P}$ \\
\hline
Total & 487 & 254 & 233 \\
\hline                                   
\end{tabular}
\\
\end{center}
$^{\P}$~This includes the 68 hard-only sources with {\it pos\_err} $> 10$\arcsec~
which were not formally included in the identification process.
\end{table}

\subsection{Counterparts of the XSS hard+soft sources} 
\label{hardsoft}

We first consider the identification statistics for the XSS sources
detected simultaneously in both the XSS hard and soft bands 
(Table \ref {table:1}, col.3).   A striking result is
potential counterparts can be identified for {\it all} of these
sources.   Rather as expected, AGN predominate,
whereas the number of clusters is small,
consistent with our earlier discussion of the
strong bias against extended sources in the XSS survey.

\begin{figure}
\begin{center}
\begin{tabular}{c}
\rotatebox{-90}{\includegraphics[height=8cm]{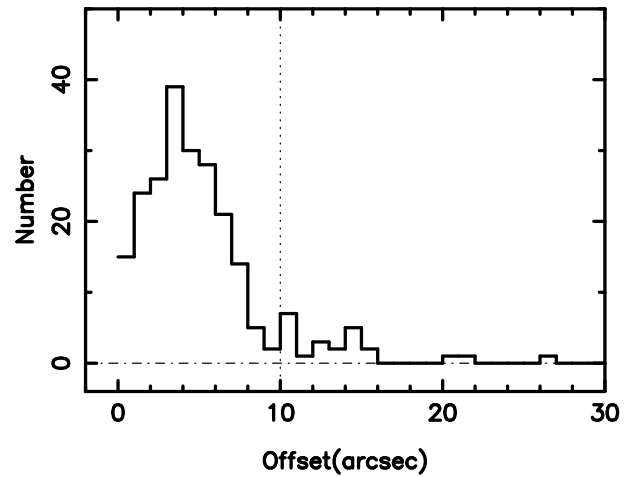}}
\end{tabular}
\end{center}
\caption
{X-ray to counterpart positional offsets for the
XSS hard+soft sources classed as either AGN, Galaxies or Stars. 
The dotted vertical line is drawn at 10\arcs~and represents a
nominal 90\% error circle radius.}
\label{fig:3} 
\end{figure}

Within the hard+soft subset, 227 sources
(out of 254) are categorized as  either {\it AGN}, {\it Galaxies} or {\it Stars}. 
Of these, 223 appear in the  WISE catalogue. For the four
remaining sources, potential identifications  were found in the GALEX catalogue
(3 sources) or in 2MASS (a bright star); hence we have a complete set of
counterpart positions for the {\it AGN}, {\it Galaxies} and {\it Stars}. 
The distribution of the angular offsets 
between the XSS position and the counterpart position 
is shown in Fig. \ref{fig:3}.  From this  distribution one can conclude that
the XSS 68\% error radius is at $\approx$ 6\arcsec~(which may be
compared to the earlier estimate of 8\arcsec~quoted by Saxton et al. 2008)
and the 90\% XSS error radius is at $\approx$ 10\arcsec.
Beyond 10\arcsec~the offset distribution exhibits a long tail
(Fig. \ref{fig:3}), which might perhaps point to some of the
identifications being incorrect.  By way of illustration, consider the
three most extreme outliers for which the putative counterpart is offset
from the XSS position by $>$ 20\arcsec. In these cases, the proposed
counterparts (1 star, 1 galaxy and 1 AGN) are all relative bright
WISE sources and are the only  candidates visible within
30\arcsec~of the X-ray position. Assuming that in these cases and 
in the others comprising the tail of the offset distribution, we do
have the correct identification, then it would appear that the
XSS position determination is occasional subject to atypical
systematic errors (of up to 20\arcsec), the origin of which are
unknown.

\begin{figure}
\begin{center}
\begin{tabular}{c}
\rotatebox{-90}{\includegraphics[height=7.5cm]{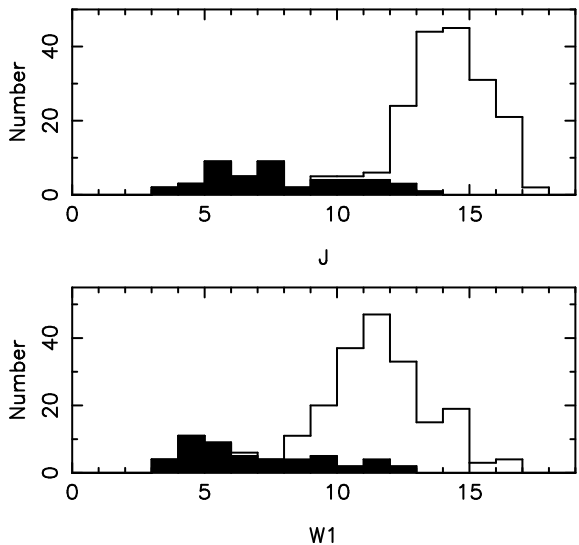}}
\end{tabular}
\end{center}
\caption
{ {\it Upper panel:} The 2MASS J magnitude distribution of
the counterparts to the XSS hard+soft sources categorized as either
{\it AGN}, {\it Galaxies} or {\it Stars}. 
{\it Lower panel:} The WISE W1 (3.4 $\mu m$) magnitude distribution
for the same source classes.
In both panels the contribution of the {\it Stars} is highlighted
as the filled histogram.}
\label{fig:4} 
\end{figure}

As noted previously, the great majority of sources classed as
either {\it AGN}, {\it Galaxies} or {\it Stars} have counterparts
visible in both the 2MASS near-IR and the WISE mid-IR
catalogues. There are in fact 213 hits with 2MASS and 223 with
WISE out of a total sample of 227 objects.  
Fig. \ref{fig:4} shows the resulting magnitude
distributions for both the 2MASS J band and the WISE W1 (3.4 $\mu m$)
band, with the contribution of objects classed as {\it Stars} highlighted.
Clearly nearby stars dominate at bright magnitudes in both the 
near- and mid-IR. 
The broad J magnitude distribution of the coronally active
stars detected in the XSS, with a shallow peak around J = 5-8, 
is in line with the much
fainter (J $\approx 13-16$) population of such objects
discovered serendipitously in pointed \xmm~observations
at X-ray flux levels $ \sim 10^{3}$ times fainter than
those reached in the XSS (see \cite{warwick11}).
The overlap region, between the
stars on the one hand and AGN/galaxies on the other, extends over
roughly 5 magnitudes in both the J and W1 bands demonstrating
that the apparent brightness of the counterpart is not a particularly
effective method  of distinguishing between Galactic and extragalactic
counterparts.  The peak in magnitude distribution
for the {\it AGN} and {\it Galaxies} is well above the
limiting sensitivity of both the 2MASS and WISE surveys, 
indicating that at the X-ray flux levels sampled by the XSS,
these two surveys provide an ideal starting point for source
identification.

\subsection{Counterparts of the XSS hard-only sources} 
\label{hardonly}

It is immediately evident from Table \ref {table:1} (col.4) that
the success rate in finding potential identifications for
the XSS sources detected solely in the hard band is very much
lower than that of sources detected simultaneously in both
the XSS bands. The large number of sources which
remain unidentified in the hard-only subset clearly
merits some investigation.

A first consideration is that the XSS hard-only
sources will have poorer position determinations than the hard+soft
by virtue of the smaller
number of counts available for the position determination
algorithm (we employ the ``best'' XSS positions which, in most cases, 
derive from all the available counts for any given source).
As noted earlier, to mitigate the impact of the poorer positions,
we excluded hard-only sources with {\it pos\_err} $>$ 10\arcsec~
from the identification statistics. The net effect of the
{\it pos\_err} constraint was to reduce the hard-only
sample from 233 sources to 165 sources. For comparison, 
if the same filter were to be  applied to the hard+soft
sources  this would lead to the removal of just 8 sources, 
5 of which lie in the {\it Clusters} category. 

\begin{figure}
\begin{center}
\begin{tabular}{c}
\rotatebox{-90}{\includegraphics[height=7cm]{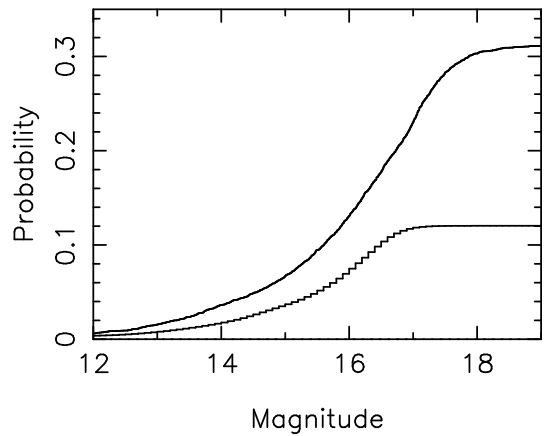}}
\end{tabular}
\end{center}
\caption
{
The probability of finding by chance an object brighter 
than a given magnitude within a 10\arcs~radius error circle. 
The upper curve corresponds to the WISE W1 (3.4 $\mu m$)
band and the lower curve to the 2MASS J band.
}
\label{fig:5} 
\end{figure}

Even with the reduced sample, a large fraction of the hard-only
sources remain unidentified (110 out of 165, {\it i.e.} 
two-thirds of the sample).
With so many unidentified sources, it was essential to take
account of the chance coincidence rate when searching
for potential counterparts down to relatively faint fluxes. 
To quantify this we investigated the probability of 
finding an object within a nominal 10\arcsec~error circle
as a function of the limiting magnitude for both
the 2MASS J and the WISE W1 bands. 
In both cases the chance rates were determined by searching for the
brightest object within a 10\arcs~circle positioned at a set of grid
points around each XSS source location (with an offset
from the actual X-ray position always $>$ 1\arcmin).
The results are shown in Fig.\ref{fig:5}.  It is evident
that at the survey limit, the WISE W1 band probes a
source density $\approx$ 2.6 times higher than that reached
in the 2MASS J band. Therefore, in the analysis below,
we use the WISE survey as our yardstick for determining the chance
coincidence rate.

Setting aside the 10 sources in the {\it Other} category, 
we conducted a systematic search for the counterparts to the
hard-only sources using the WISE survey. 
For putative counterparts brighter than W1 = 12 
we extended the search radius to 20\arcsec,
whereas for objects in the range W1 = 12--15 we restricted
the radius to 10\arcsec~(noting that the latter equates 
to the 90\% error radius for the hard+soft sources). 
Candidate objects fainter than W1 = 15 were not considered. 
The number of potential counterparts as a function of the W1
magnitude emerging from this process is listed in Table \ref{table:2},
together with an estimate of the number of source expected
by chance across the 165 sources comprising the (reduced) hard-band only
sample. In total 45 potential counterparts were found - as reported
in Table \ref {table:1}. The corresponding number of chances
coincidences was predicted to be 10.6. We note from Table \ref{table:2}
that at the faint end of the magnitude range considered, 
the chance rate is rapidly approaching the actual number of objects
found. We conclude, that a limiting W1 magnitude of
$\approx 15$ and a search radius at faint fluxes of no more than 
10 \arcsec~defines a reasonable bound to counterpart searches for the
hard-only XSS sample.

\begin{table}
\caption{Incidence of potential WISE counterparts
amongst the 165 sources comprising the (reduced) hard-only sample
as a function of the W1 (3.4 $\mu m$) magnitude.}
\label{table:2}
\begin{center}
\begin{tabular}{l c c c}   
\hline
W1    & Search Radius  &  No. of      &  Chance \\
magnitude         & (arcsec)       & Counterparts   &  Rate   \\
\hline
$<$10      & 20 & 2 & 1.0  \\
10-11    & 20 & 9 & 1.2  \\
11-12    & 20 & 17 & 1.4 \\ 
12-13    & 10 & 5 &  1.2 \\ 
13-14    & 10 & 7 &  2.4 \\
14-15    & 10 & 5 &  3.4 \\
\hline
Total    &   & 45 & 10.6 \\
\hline
\end{tabular}
\\
\end{center}
\end{table}

Of the 45 hard-only sources classed as {\it AGN}, {\it Galaxies}
or {\it Stars}, only 2 fall in the latter category. This presumably reflects
the fact that coronally active stars generally have soft thermal spectra and 
are not subject to significant line-of-sight absorption (particularly
when detected at high Galactic latitude). The split between {\it AGN} and
{\it Galaxies}
is also much more strongly weighted to the latter in the hard-only sample. 
This could be the result of
absorbing material within the host galaxy suppressing both 
the soft X-ray emission and also masking the AGN character of
the source at longer wavelengths.   

This leaves us with some uncertainty as to the nature of the
numerous unidentified sources in the hard-only sample.
If one applies the same filtering to the hard+soft identification as
applied to the hard-only sources ({\it pos\_err} $<$ 10\arcsec,
W1 $<15$, search radii as in Table \ref{table:2}), then the number
of sources ({\it AGN}, {\it Galaxies} and {\it Stars}) with putative WISE
counterparts drops from 233 to 205, {\it i.e.,} a modest 12\% decrease.
It would seem that the unidentified sources amongst the hard-only
sample, if real, must represent a population of astrophysical sources 
substantially fainter than found amongst the XSS hard+soft subset. 
However, there is no real evidence for the emergence of
such a population down to W1=15, which is 3 magnitudes fainter
than the peak of distribution in Fig.\ref{fig:4}.  One might speculate
that the unidentified sources correspond to infrequent, relatively short-lived
outbursts linked to an otherwise underluminous source population, but in
truth they are difficult to match to any known class of X-ray
emitting object (see also \S \ref{source_counts}).  Unfortunately,
a follow-up study of a subset of XSS sources with SWIFT provided
no information contrary to the above analysis (\cite{starling11}).

The alternative possibility is that a great many of the unidentified sources
are not real astrophysical sources, but rather false detections, mostly
at the 4-count threshold. The argument against this hypothesis is that the numbers
of sources involved exceed the estimates of \S\ref{false} by a factor of at least 3.
However, given the nature of the XSS, a higher than predicted false detection
rate is certainly a possibility. At the very least some caution is
needed when dealing with XSS sources detected only in a single band
at the 4-count limit.

\section{The XSS extragalactic sample}

\subsection{Composition of the sample}
\label{sample} 

We may now use the source identifications to construct an XSS
extragalactic sample. To this end, we have included the sources classed
as either {\it AGN} or {\it Galaxies} in  Table \ref {table:1},
but excluded {\it Clusters} given the strong bias against
such sources in the XSS. Our full XSS extragalactic sample comprises
219 sources.  Details of the individual sources are provided in the
appendix (Table\ref{table:cat}), where we list the counterpart
position, source type and redshift (if known).

The  composition  of the extragalactic sample is presented in
Table~\ref{table:3} for the full set of sources and also split 
between the hard+soft and hard-only subsets.  Here the AGN
are grouped according to their spectroscopic classification,
as taken from the literature. As noted earlier
a number of sources initially categorized as
{\it Galaxies} were switched to {\it AGN} on the basis
of their UV to near-IR colour. The specific requirement was
for NUV - J $<$ 5, where NUV is the near-UV magnitude from GALEX
and the J magnitude is from the 2MASS.  The 18 objects that met
this criterion are classed as UVX sources both in
Table \ref{table:3} and Table~\ref{table:cat}.

\begin{table}
\caption{Division into source types in the XSS extragalactic sample}
\label{table:3}
\begin{center}
\begin{tabular}{l r r r}   
\hline
Type    & Full   &  Hard+soft &  Hard-only \\
        & Sample &  Subset   & Subset \\
\hline
Liner       &  2 &  2 &  0 \\
Sy 1-1.5    & 79 & 73 &  6 \\
Sy 1.9-2    & 14 &  7 &  7 \\
RG/QSO      & 40 & 35 &  5 \\
BLLac       & 28 & 27 &  1 \\
UVX         & 18 & 16 &  2 \\
Galaxy      & 38 & 16 & 22 \\
\hline
Total      & 219 & 176 & 43 \\
\hline
\end{tabular}
\\
\end{center}
\end{table}

\begin{figure}
\begin{center}
\begin{tabular}{c}
\rotatebox{-90}{\includegraphics[height=8.5cm]{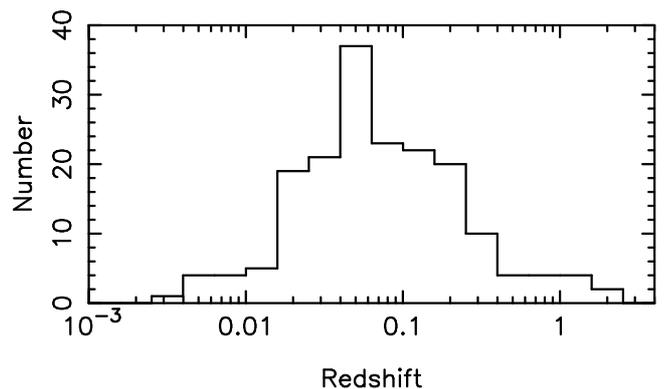}}
\end{tabular}
\end{center}
\caption
{Redshift distribution of the sources in the XSS extragalactic sample.
Redshifts are known for 181 out of the 219 sources which comprise the sample.
}
\label{fig:6} 
\end{figure}

Of the 219 AGN and galaxies comprising the XSS extragalactic sample, 181
have spectroscopic or photometrically estimated redshifts 
(see Table \ref{table:cat}). 
The distribution of redshifts is shown in Fig. \ref{fig:6}.
The mode of the distribution is at z $\approx 0.05$; for a source at the survey limit
the corresponding X-ray luminosity (for $H_{0} = 70 \rm~km~s^{-1}~Mpc^{-1}$)
is $\sim 1.7 \times 10^{43} \rm~erg~s^{-1}$ (2--10 keV). 

\subsection{Infrared colours}
\label{multiwavelength}

\begin{figure}
\begin{center}
\begin{tabular}{ccc}
\rotatebox{-90}{\includegraphics[height=7cm]{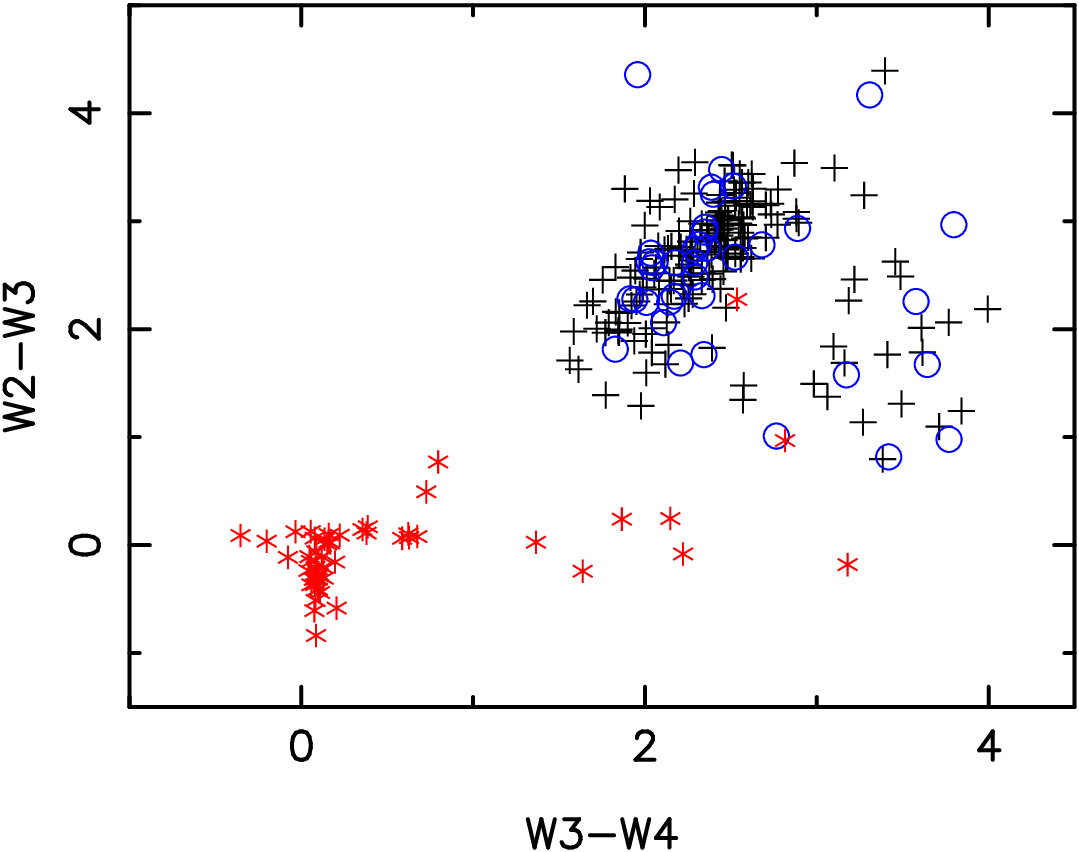}}\\
\rotatebox{-90}{\includegraphics[height=7cm]{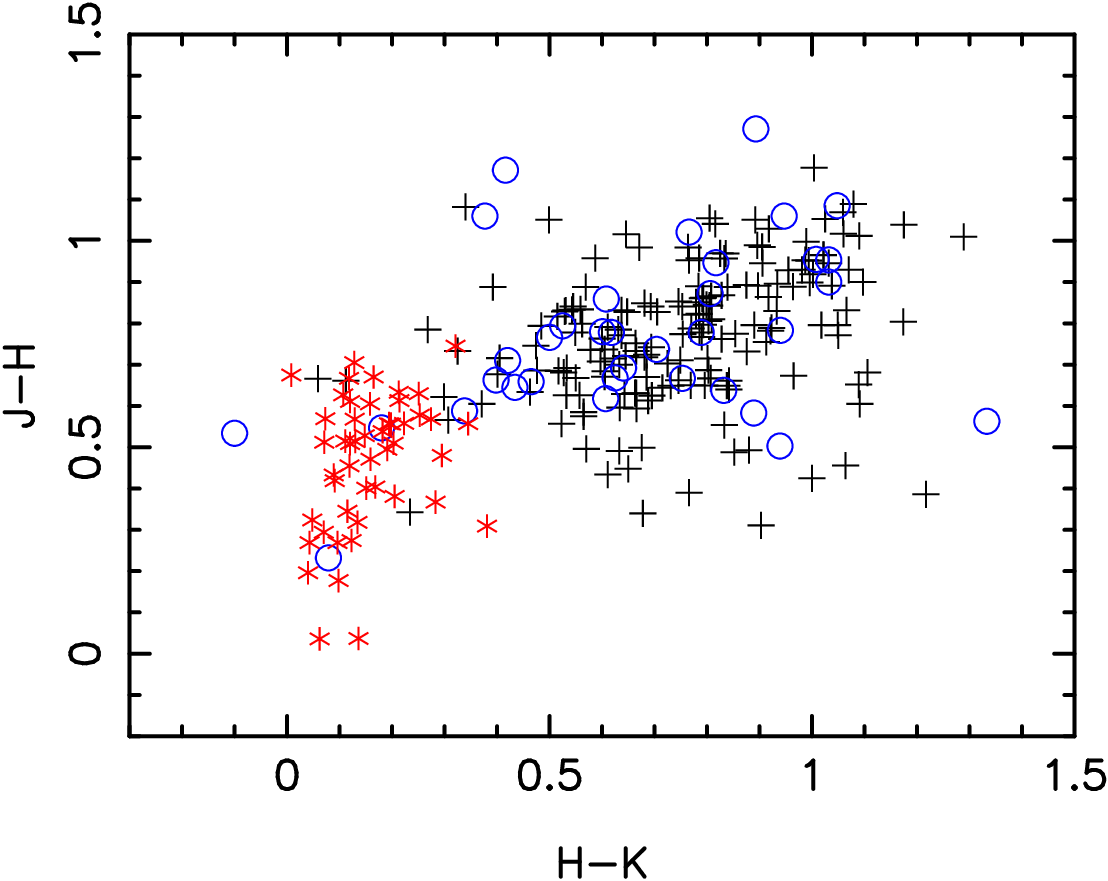}}\\
\end{tabular}
\end{center}
\caption
{The mid- and near-IR colours of the AGN and galaxies in the
XSS extragalactic sample.
{\it Top panel:} WISE  W2-W3 versus  W3-W4 two-colour diagram.
{\it Bottom panel:} 2MASS   J-H versus H-K two-colour diagram.
In both cases, the XSS AGN are shown as crosses (black) and the galaxies as
circles (blue). For comparison, the XSS sources categorized as
stars are shown as asterisks (red).
}
\label{fig:7} 
\end{figure}

We have investigated the mid-IR and near-IR colours of the AGN
and galaxies in the XSS extragalactic sample by utilising
the WISE 3.4, 4.6, 12 and 22 $\mu m$ data ({\it i.e.,} the W1, W2, W3
and W4 bands respectively), and the 2MASS J, H and K bands.
Mid-IR and near-IR two-colour diagrams are shown in Fig. \ref{fig:7}.
These diagrams demonstrate that the AGN and galaxies are well separated from the stellar
coronal sources in terms of their W2-W3 colour (apart from one or two early type
stars found in local star formation regions). With slightly less
efficiency, the H-K colour also serves to distinguish between the extragalactic
and Galactic X-ray source populations.
We note that the sources in the XSS extragalactic sample occupy largely the same
region of the mid-IR two-colour diagram as the luminous AGN in the 
Bright Ultra-Hard {\it XMM-Newton} Survey (\cite{mateos12}).

\subsection{Cross-correlation with the SWIFT BAT catalogue} 

\begin{figure}
\begin{center}
\begin{tabular}{c}
\rotatebox{-90}{\includegraphics[height=7cm]{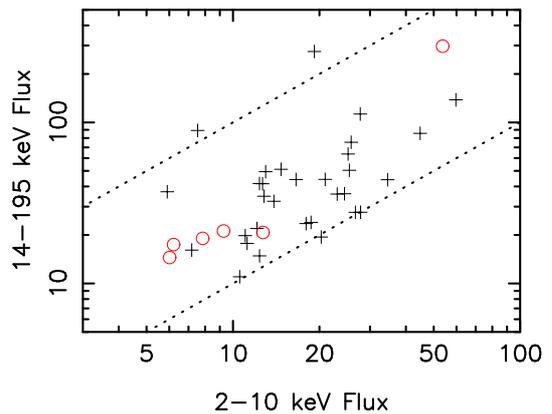}}
\end{tabular}
\end{center}
\caption
{The correlation between the 14-195 keV flux (10$^{-12}$ erg~cm$^{-2}$~s$^{-1}$) measured
  by the BAT experiment on Swift and the XSS 2--10 keV flux 
  (10$^{-12}$ erg~cm$^{-2}$~s$^{-1}$) for the 37 cross matches.  XSS hard+soft are shown
  as crosses, whereas XSS hard-only sources are shown as circles (red).  The
  dotted lines represent BAT:XSS flux ratios of 1:1 and 10:1 respectively.  }
\label{fig:8} 
\end{figure}

We have also investigated the coincidence of the XSS extragalactic sample
with sources discovered in the hard (14-195 keV) X-ray band by 
the BAT experiment on Swift.
For this purpose we use the on-line\footnote{Available at
http://heasarc.nasa.gov/docs/swift/results/bs58mon/.}
Swift BAT 58-Month Hard X-ray Survey (\cite{baumgartner10}).
Using a 1\arcm~search radius, there were 37 XSS-BAT matches,
6 of which were with XSS hard-only sources. 
Fig. \ref{fig:8} shows the correlation diagram for the 14-195 keV
versus 2--10 keV fluxes, in which diagonal dotted lines
delineate 14-195 keV:2-10 keV flux ratios of 1:1 and 10:1.
For comparison, the lower ratio applies to a source exhibiting
relatively soft continuum emission ($\Gamma \sim 2$) 
subject to a modest level of absorption ($N_H \approx 10^{21} \rm~cm^{-2}$),
whereas a requisite for the higher ratio is either a very hard continuum 
($\Gamma \sim 1.0$) or substantial line-of-sight absorption 
($N_H > 10^{23} \rm~cm^{-2}$).
 
Interestingly there were no BAT detections of any of the XSS sources
classed as {\it Unidentified}. 

\section{Determination of the XSS source counts}

A major objective of this investigation was to establish the log N - log S 
relation for a sample of extragalactic sources selected in the
hard 2--10 keV from the XSS  and to compare the results with 
those reported from other surveys. Given the small numbers of clusters 
detected in the XSS, the focus is necessarily on the identified sample
of AGN and galaxies.

\subsection{The results from a Monte Carlo simulation}

Given the wide range of factors which might influence the detection probability
of {\it real} sources and, for those sources with positive detections,
might deviate the measured count rate from the true value, we have carried
out a Monte Carlo simulation of the survey process. 

In brief, this involved adding simulated sources at random positions
to the set of sky images created as part of the slew survey pipeline.
For a simulated source of specified count rate,  the number of net counts 
was calculated on the basis of the effective exposure and then 
randomised according to Poissonian statistics. These counts 
were then distributed about the assigned source position according to
the appropriate point spread function. The source detection algorithm was then 
applied and the number of counts net of the background determined
from which a {\it measured count rate} could be assigned.
The simulated source was classed as a detection and considered further provided all the criteria
applied in the actual XSS selection were met (\eg a minimum of 4 net counts).  
Repetition of this process then allowed a count rate probability distribution
appropriate to the specified input rate to be determined. Finally a series of 
trials were conducted so as to cover a representative range of input 
count rate; each of these trials were based on 10000 simulated sources, 
except for the two trials at the lowest input count rates (0.1 and 0.2 ct/s) where 20000
sources were employed and the trial with the highest input rate (10 ct/s)
which was based on 4000 sources.

\begin{figure}
\begin{center}
\begin{tabular}{c}
\rotatebox{-90}{\includegraphics[height=8cm]{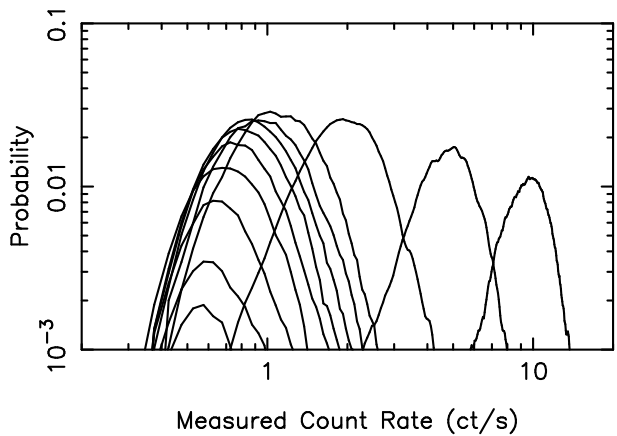}}
\end{tabular}
\end{center}
\caption
{Probability of measuring a given output count rate for simulated sources
of fixed input count rate.  These probability density functions are
based on bins of 0.05 ct/s width and are smoothed versions
of the raw simulation data (using a 1-d top-hat smoothing filter
with a width of 5 bins). From left to right (based on the
position of the peak) the 
curves correspond to input count rates  of 
0.1, 0.2, 0.3, 0.4, 0.5, 0.6, 0.7, 
0.8, 1.0, 2.0, 5.0 and 10 ct/s.  For this plot
the normalisation of the 0.1 ct/s curve
has been scaled upwards by a factor of 5.
}
\label{fig:9} 
\end{figure}

Figure \ref{fig:9} shows the results of the simulation in the form of 
probability density curves plotted as a function of the measured count rate.  
The set of curves cover two decades of input count 
rate from 0.1 ct/s up to 10 ct/s.
The detection probability 
evidently truncates at a measured count rate of $\sim 0.4$ ct/s, 
\ie the 4 count threshold divided by the maximum exposure of $\approx 10$ s.  
We note that even at very low input rates (\eg 0.2 ct/s for which the 
average count at maximum exposure is 2),  there is 
still a finite probability that the source will be detected,
since Poissonian deviations allow the occasional 
detection of the source at or above the 4 count threshold.

\begin{figure}
\begin{center}
\begin{tabular}{c}
\rotatebox{-90}{\includegraphics[height=7cm]{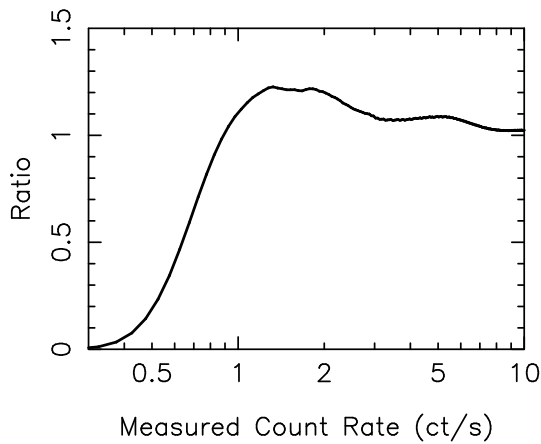}}
\end{tabular}
\end{center}
\caption
{Ratio of the predicted differential log N - log S function
(as determined by the simulation) to the assumed input power-law
form.}
\label{fig:10} 
\end{figure}

The next step was to use the curves in
Fig. \ref{fig:9} to predict the number of sources detected as a function
of the measured count rate.  Here we took the underlying differential 
log N - log S function of the cosmic X-ray source population to 
be a powerlaw function with index $\gamma$ and normalisation $K$:

\begin{equation}
N(S)dS  =  K S^{-\gamma}dS
\end{equation}

where $S$ is the true count rate in ct/s units and $N(S)dS$ is the
number of sources with true count rate in the range $S$ to $S+dS$.
We assumed $\gamma = 2.5$ (\ie the Euclidean form of the source counts).


The range $S$ = 0.1 - 20 ct/s was divided into bins (of width
0.01 ct/s for S = 0.10--0.25 ct/s and 0.1 ct/s thereafter) and 
the number of sources $\Delta N$ within the bin calculated.  For each
value of $S$ the corresponding  probability density $P(S_m)$ 
as a function of the measured count rate, $S_m$ was determined, where necessary by
interpolating across the curves shown in Fig.\ref{fig:9}.   The
summation of the $\Delta N \times P(S_m)$ products over the full range of
$S$ then allowed the differential source count as a function
of the {\it measured count rate}, $M(S_m)$, to be determined.

The ratio $M(S_{m}))/N(S_m)$ is then a prediction,
based on the simulation, of how closely the measured source counts
will track the input form.  Figure \ref{fig:10} shows this ratio for the
range of count rates encompassed by our XSS sample. The simulation demonstrates 
the onset of a cutoff below a measured count rate of 1 ct/s with a sharp 
truncation at $\approx  0.4$ ct/s. The prediction of a slight excess of sources 
with measured count rates from 1--3 ct/s most likely stems from the effect of
Eddington bias.  This is the process by which, through Poissonian fluctuations,
faint sources are boosted to higher levels more frequently than bright sources 
are suppressed, by virtue of the numerical superiority of the former.

\begin{figure}
\begin{center}
\begin{tabular}{c}
\rotatebox{-90}{\includegraphics[height=8.8cm]{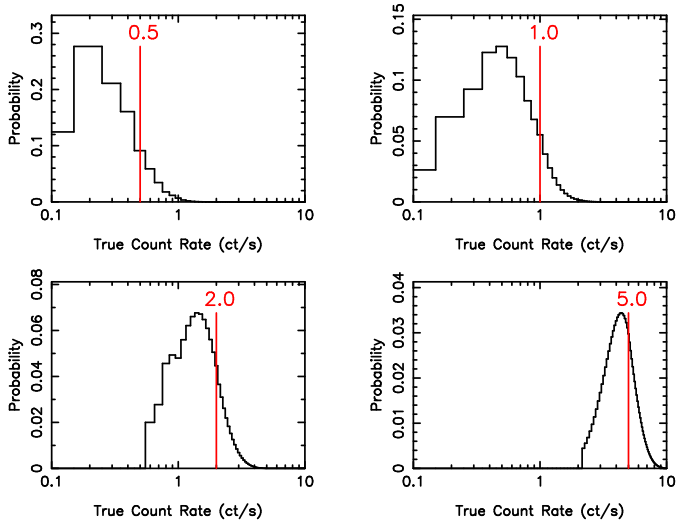}}
\end{tabular}
\end{center}
\caption
{Spread in the true (input) count rate for sources detected at different
{\it measured} count rates.  The four panels show the results for measured count rates
of 0.5, 1.0, 2.0 and 5.0 ct/s (as represented by the vertical lines).
}
\label{fig:11} 
\end{figure}

Further evidence of the major impact of Eddington bias on the XSS is provided in
Fig. \ref{fig:11} which shows the spread in the true (input) count rate
for sources detected at a particular measured count rate.  For example, 
a measured count rate of 0.5 ct/s can derive from sources with true count
rate anywhere between 0.1-- 1 ct/s. Even for much brighter sources the
underlying uncertainty in the true count rate (and hence the source
flux) is surprisingly large.

\subsection{Extragalactic log N - log S relation}
\label{source_counts}

We have determined the source counts of the set of sources which comprise our
extragalactic sample. A technical issue arises in relation to how to deal with the overlap of the
{\it XMM-Newton} slews which comprise the XSS.  As noted earlier, this overlap
results in some duplicate source detections in the primary XSS database, which we
subsequently removed in building our hard-band selected sample.  However,
the source count calculation is greatly simplified if we ignore this overlapping
coverage, but commensurately add back the duplicate detections. In practice
this involved adding 17 duplication detections to the set of 219 sources which
comprise our extragalctic sample.

\begin{figure}
\begin{center}
\begin{tabular}{cc}
\rotatebox{-90}{\includegraphics[height=7.5cm]{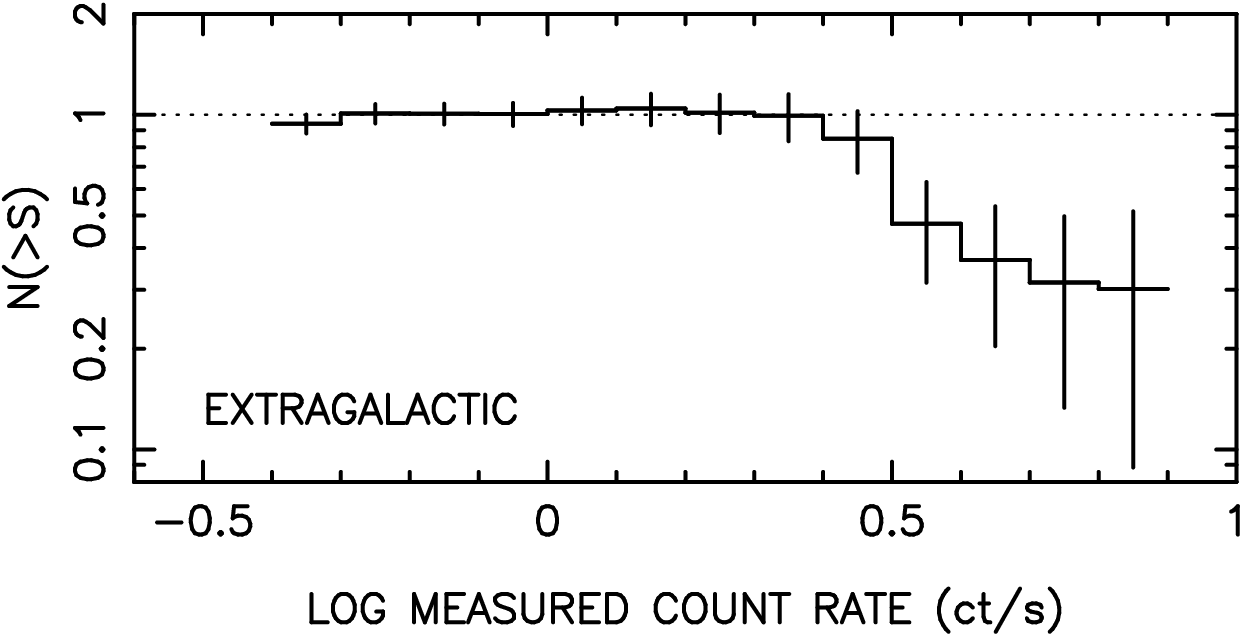}}\\
\rotatebox{-90}{\includegraphics[height=7.5cm]{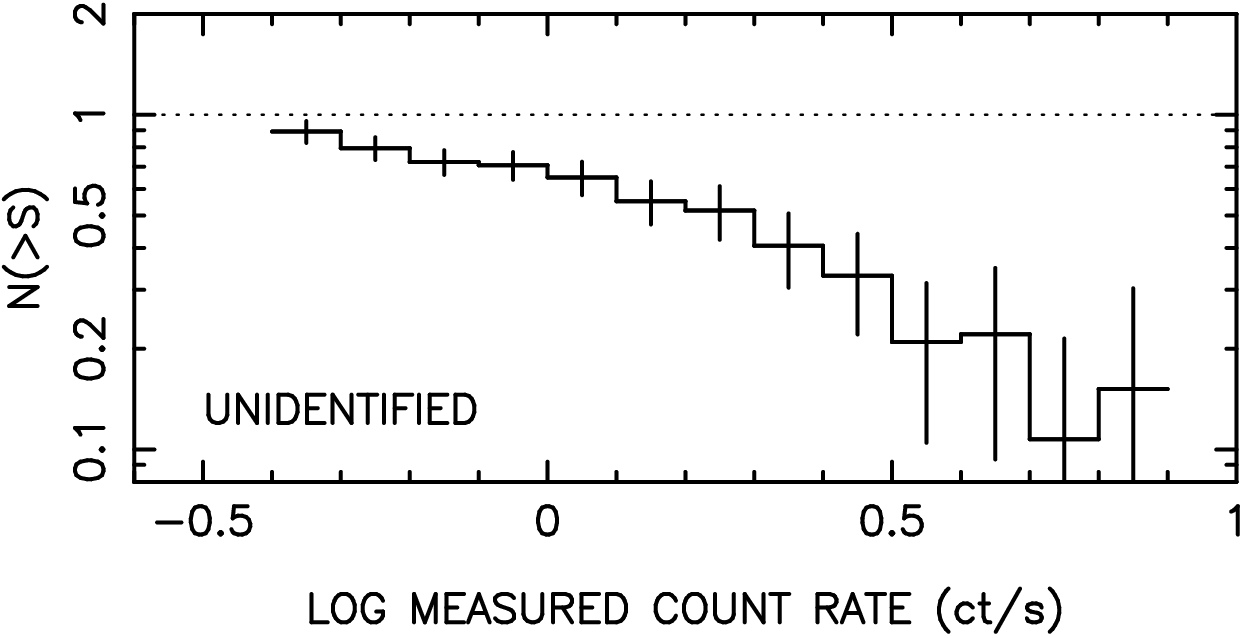}}\\
\end{tabular}
\end{center}
\caption
{{\it Top panel:} Integral log N - log S curve for the extragalactic XSS sample
(excluding clusters of galaxies) normalised to $N(>S)=100 \times S^{-1.5}$.
{\it Bottom panel:} Integral log N - log S curve for the sources
classed as {\it Unidentified} in the XSS hard-band selected sample.
}
\label{fig:12} 
\end{figure}

The integral source counts were determined by giving each source a weight 
$N(S_m)/M(S_m)$ (\ie the reciprocal of the function
plotted in Fig. \ref{fig:10}) and then summing these
weights as a function of decreasing $S_m$.
The result is shown in Fig. \ref{fig:12} in a normalised form, \ie the measured
integral counts divided by the integral version of the fiducial Euclidean
source count defined previously.  

The integral log N - log S curve for the extragalactic sources follows the Euclidean
form below 3.0 ct/s.  The normalisation implies 100 sources brighter than 1 ct/s
are detected in the 14233 square degree covered by the XSS (overlaps included),
which, down to the limiting sensitivity of the XSS (0.4 ct/s $ = 
3.2 \times 10^{-12} \rm~erg~cm^{-2}~s^{-1}$) implies $\sim 1000$ X-ray sources
in the high-latitude sky above $|b| > 10^{\circ}$
(excluding, of course, clusters of galaxies and Galactic objects).
In Fig. \ref{fig:12} there is an  apparent deficit of sources above 3.0 ct/s.  
In fact using the normalisation quoted above and assuming the Euclidean
form for the counts extends from 3--10 ct/s, we predict 14 sources
in this count rate range compared to the 9 sources in the actual
sample. For a Poisson distribution, the chances of
recording 9 or less events for a mean rate of 14 is 11\%, implying that
the deficit of bright sources in the XSS is not hugely significant.

By way of comparison, Fig. \ref{fig:12} also shows the integral counts for
the 178 sources classed as {\it Unidentified} in the XSS hard-band selected sample.
The steep form of this relation ($\gamma \approx 3.0$) is difficult to match to
any known astrophysical population and hence adds weight to the conjecture that
most of these sources are spurious.

\begin{figure}
\begin{center}
\begin{tabular}{c}
\rotatebox{-90}{\includegraphics[height=8cm]{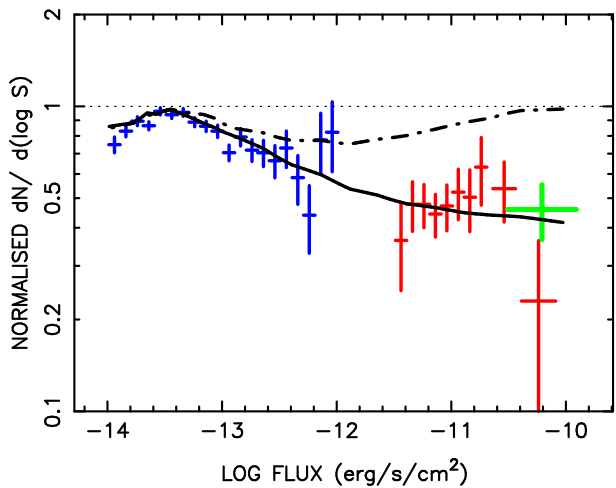}}
\end{tabular}
\end{center}
\caption
{Differential 2--10 keV log N - log S curve for the XSS extragalactic
sample (red data points), normalised to the 
Euclidean form.  At the faint end the XSS results are compared with
the differential 2--10 keV source counts derived from  the 2XMM catalogue
by Mateos et al. (2008) (blue points).  At the bright 
end the comparison is with the {\it HEAO-1} A2 AGN sample of Piccinotti et al. 
(1982) (green point). The solid curve is the prediction of the AGN source 
counts from the work of Gilli, Comastri \& Hasinger (2007).  The dot-dashed 
curve similarly represents the prediction of the total extragalactic counts,
\ie AGN plus clusters of galaxies.
In drawing these comparisons account has been taken of the 
relevant survey sky areas. In this plot the normalisation factor has been
matched to the counts of the AGN plus clusters at the high flux end. 
}
\label{fig:13} 
\end{figure}

We have also derived the {\it differential} form of the
log N - log S curve for the extragalactic sample.
Figure \ref{fig:13} shows the result, again normalised to the
Euclidean form. So as to make comparisons with other surveys,
we have converted the XSS count rates to 2--10 keV fluxes 
using the conversion factor specified in \S 2.1. As noted above,
within the flux range encompassed by the XSS, the counts 
of extragalactic objects (excluding clusters) follow the 
Euclidean form.
Figure \ref{fig:13} also compares the XSS results with the 
differential 2--10 keV source counts at high latitude
derived from the 2XMM catalogue (\cite{mateos08}).
The normalisation of the source counts at the high
flux end of the 2XMM survey and at the low flux end of the XSS
are very comparable, suggesting that a smooth interpolation
is possible across the intervening gap (which amounts to a factor 
$\sim 3$ in flux). There is, however, the caveat that
the 2XMM data do not specifically exclude either
Galactic interlopers or clusters of galaxies.
Figure \ref{fig:13} also shows a point at high flux
corresponding to the AGN sample defined by Piccinotti et al. (1982)
on the basis of the {\it HEAO-1} A2 survey.  
Here the agreement with XSS is good, apart from the apparent
deficit in the highest XSS bin\footnote{In this case roughly 4 
additional sources would be required in the highest flux bin 
(encompassing sources with count rates from 5-10 ct/s) 
so as to raise this bin to an ordinate value of 0.5.}.

To aid the interpretation of the source counts we have also plotted in Fig. \ref{fig:13},
the predictions of  Gilli, Comastri \& Hasinger (2007) for the 2--10 keV band both for
AGN alone and for AGN plus clusters of galaxies. There is clearly excellent agreement
between the AGN prediction and the measured XSS counts. In the flux range of
the XSS, we might expect that the inclusion of clusters
in the extragalactic sample would result in an additional  80\%-100\% 
of sources (consistent with the Piccinotti et al results where
clusters represented half the total extragalactic sample).  Clearly 
this reemphasises the point made earlier that there is a very strong selection
bias against extended sources in the XSS catalogue.
Figure \ref{fig:13} also provides at least a hint that some clusters of galaxies
may also be missing from the 2XMM catalogue (\cite{mateos08}).

\section{Summary and Conclusions}

The substantial sky coverage afforded by the XSS makes this survey a 
unique resource for studying X-ray bright source samples. 
We have shown that, with care, the XSS can be used to define reliable
subsets of sources selected in the 2--10 keV band, down to a limiting sensitivity
of $3 \times 10^{-12} \rm~ergs~cm^{-2}~s^{-1}$ and encompassing sources with
as few as 4 counts net of the background.  One caveat is that there is a
significant bias against the detection of extended sources, such as clusters
of galaxies.  A second caveat is that a significant fraction ($\approx$ 75\%) of
the faint XSS sources detected solely in the hard band may in fact be spurious
detections.

For the purpose of source identification, the XSS is well matched to currently
available all-sky IR surveys such as 2MASS and WISE. For the XSS sources 
detected simultaneously in the XSS hard and soft bands, the hit rate with
WISE was close to 100\%, in the case of the AGN/galaxies and stellar coronal
emitters. The infrared colours of the counterparts also provides a rather
efficient method of removing the active stars when constructing an
extragalactic sample. Unfortunately, for the XSS sources detected 
only in the hard band, the search for WISE counterparts is somewhat compromised
by the high chance rate that applies when the sample size is inflated by spurious
detections.

A major motivation of this work was to use the XSS to determine the extragalactic
log N - log S relation in a flux regime which is poorly sampled by existing surveys. 
To that end we  constructed an XSS extragalactic sample comprised of 219 
sources with likely identifications as AGN/galaxies, but necessarily
excluding clusters of galaxies.  Using the results of a detailed simulation
of the source detection procedure we were able to apply corrections for the complex
selection biases which come into play in the low-count regime of the XSS.
The conclusion of this study was that the normalisation we derive for
the XSS extragalactic source counts fits well with published measurements
at brighter and fainter levels, which together span 4 decades in X-ray
flux.

\acknowledgements

AMR acknowledges STFC/UKSA funding support.
We thank Silvia Mateos for providing the 2XMM source count data
in a convenient form. This research has made use of the
SIMBAD and Vizier facilities at CDS, Strasbourg and the NASA/IPAC 
Extragalactic Database (NED). This publication utilises data
from  the Two Micron All Sky Survey, which is a joint project of the
University of Massachusetts and IPAC/Caltech. This paper has also made
use of the source catalogue from the Wide-field Infrared Survey Explorer,
which is a NASA sponsored joint project of the University of California,
Los Angeles, and the Jet Propulsion Laboratory/California Institute of
Technology. Use was also made of data from the Galaxy Evolution
Explorer, GALEX, which is operated for NASA by the California
Institute of Technology under NASA contract NAS5-98034.
The XMM-Newton project is an ESA science mission with instruments and
contributions directly funded by ESA member states and the USA (NASA).

\newpage

\appendix

\section{}

\normalsize

Details of the sources which comprise the hard-band selected XSS extragalactic
sample - see Table \ref{table:cat}. The table provides the following
information for each source: the XSS name; whether the source was also
detected in the XSS soft band; the XSS hard band (2$-$10\,keV) flux and error
on the flux (in units of $10^{-11} \rm~ergs~cm^{-2}~s^{-1}$) ; the RA and Dec
of the proposed counterpart; the name of the counterpart; the type of the
counterpart; the redshift (if known). The table is available as a FITS file
from the authors.

\vspace{1cm}
\begin{table*}
\caption{Sources comprising the XSS extragalactic sample}
\label{table:cat}
\begin{center}                    
\begin{tabular}{llrrrllr}   
\hline
Slew ID & Soft & Flux & RA & Dec & ID & Type & Redshift \\ 
\hline

XMMSL1 J000226.3+032108 & Y & 0.64$\pm$0.22 & 0.6101 & 3.3520 & NGC 7811 & Sy1 & 0.025 \\
XMMSL1 J001439.7+183449 & - & 0.56$\pm$0.18 & 3.6671 & 18.5822 & NGC 52 & Gal & 0.018 \\
XMMSL1 J003621.2+453958 & Y & 0.72$\pm$0.21 & 9.0873 & 45.6649 & Zw 535.012 & Sy1.2 & 0.047 \\
XMMSL1 J004146.6-470141 & Y & 0.77$\pm$0.19 & 10.4459 & -47.0269 & RBS 97 & BLLac & 0.15 \\
XMMSL1 J004755.6+394907 & Y & 0.56$\pm$0.21 & 11.9801 & 39.8160 & 5C 3.178 & BLLac & 0.252 \\
XMMSL1 J004819.1+394119 & Y & 0.42$\pm$0.20 & 12.0791 & 39.6866 & B0045+3926 & Sy1 & 0.134 \\
XMMSL1 J005953.1+314934 & Y & 1.43$\pm$0.32 & 14.9720 & 31.8270 & Mrk 0352 & Sy1 & 0.014 \\
XMMSL1 J010512.2+802712 & Y & 0.60$\pm$0.17 & 16.2970 & 80.4520 & WISE J010511.28+802707.2 & Gal & - \\
XMMSL1 J011209.0+330310 & Y & 0.45$\pm$0.21 & 18.0400 & 33.0538 & GALEX J011209.5+330313 & UVX & 0.159 \\
XMMSL1 J011423.7-552358 & - & 0.78$\pm$0.24 & 18.6039 & -55.3970 & NGC 454E & Sy2 & 0.012 \\
XMMSL1 J012345.1-584822 & Y & 2.54$\pm$0.65 & 20.9407 & -58.8058 & FAIRALL 9 & Sy1 & 0.047 \\
XMMSL1 J012950.7-421934 & - & 1.68$\pm$0.52 & 22.4631 & -42.3265 & ESO 244-30 & QSO & 0.025 \\
XMMSL1 J015025.1+091453 & Y & 0.55$\pm$0.20 & 27.6056 & 9.2476 & PHL 1186 & Sy1.5 & 0.27 \\
XMMSL1 J015251.0-523830 & Y & 0.42$\pm$0.19 & 28.2149 & -52.6401 & GALEX J015251.3-523826 & UVX & 0.077 \\
XMMSL1 J020124.1+021545 & - & 0.91$\pm$0.39 & 30.3518 & 2.2643 & WISE J020124.44+021551.3 & Gal & 0.085 \\
XMMSL1 J020306.6-385423 & Y & 0.49$\pm$0.18 & 30.7782 & -38.9065 & PGC 109903 & RG & 0.144 \\
XMMSL1 J020615.7-001731 & Y & 1.39$\pm$0.36 & 31.5666 & -0.2914 & Mrk 1018 & Sy1.5 & 0.042 \\
XMMSL1 J020921.4-522922 & Y & 2.11$\pm$0.53 & 32.3400 & -52.4897 & RBS 0285 & BLLac & - \\
XMMSL1 J021734.8-725126 & Y & 0.91$\pm$0.25 & 34.3986 & -72.8578 & SUMSS J021735-725125 & UVX & 0.29 \\
XMMSL1 J022625.4-282054 & Y & 0.73$\pm$0.26 & 36.6071 & -28.3497 & HE 0224-2834 & Sy1 & 0.06 \\
XMMSL1 J023005.6-085950 & Y & 1.34$\pm$0.38 & 37.5230 & -8.9981 & Mrk 1044 & Sy1 & 0.016 \\
XMMSL1 J023248.4+201720 & Y & 1.49$\pm$0.33 & 38.2025 & 20.2881 & BWE 0229+2004 & BLLac & 0.14 \\
XMMSL1 J023819.5-521129 & Y & 1.90$\pm$0.42 & 39.5821 & -52.1923 & ESO 198-24 & Sy1 & 0.045 \\
XMMSL1 J030315.6-121326 & Y & 0.80$\pm$0.21 & 45.8149 & -12.2249 & PGC 955931 & RG & 0.077 \\
XMMSL1 J031028.1-004950 & Y & 0.37$\pm$0.15 & 47.6159 & -0.8307 & LBQS 0307-0101 & Sy1 & 0.08 \\
XMMSL1 J031118.7-204618 & Y & 2.78$\pm$0.63 & 47.8284 & -20.7717 & HE 0309-2057 & Sy1 & 0.066 \\
XMMSL1 J035141.4-402758 & Y & 1.06$\pm$0.31 & 57.9237 & -40.4665 & FAIRALL 1116 & Sy1 & 0.058 \\
XMMSL1 J035257.3-683117 & Y & 3.62$\pm$0.64 & 58.2395 & -68.5214 & PKS 0352-686 & RG & 0.087 \\
XMMSL1 J035324.7-404927 & - & 0.82$\pm$0.35 & 58.3533 & -40.8265 & WISE J035324.78-404935.2 & Gal & - \\
XMMSL1 J041241.2-471252 & Y & 0.54$\pm$0.17 & 63.1728 & -47.2128 & RBS 518 & Sy1 & 0.132 \\
XMMSL1 J041959.8-545618 & Y & 1.25$\pm$0.43 & 65.0016 & -54.9380 & NGC 1566 & Sy1.5 & 0.0050 \\
XMMSL1 J042224.4-561333 & - & 0.51$\pm$0.20 & 65.6002 & -56.2256 & ESO 157-23 & Sy2 & 0.043 \\
XMMSL1 J042559.7-571142 & Y & 1.80$\pm$0.30 & 66.5030 & -57.2005 & 1H0419-577 & Sy1 & 0.104 \\
XMMSL1 J043429.6+712747 & Y & 0.45$\pm$0.16 & 68.6213 & 71.4672 & IRAS 04288+7121 & Sy1 & 0.024 \\
XMMSL1 J043958.6-594049 & - & 0.77$\pm$0.24 & 69.9960 & -59.6816 & ESO 118-33 & Sy2 & 0.057 \\
XMMSL1 J044347.0+285822 & Y & 1.30$\pm$0.47 & 70.9450 & 28.9719 & UGC 3142 & Sy1 & 0.021 \\
XMMSL1 J045003.8+260016 & - & 0.91$\pm$0.41 & 72.5175 & 26.0049 & WISE J045004.20+260017.7 & Gal & 0.373 \\
XMMSL1 J045014.5-142557 & - & 0.40$\pm$0.15 & 72.5618 & -14.4326 & IRAS 04479-1431 & Gal & 0.038 \\
XMMSL1 J050432.4-734925 & - & 0.69$\pm$0.20 & 76.1428 & -73.8243 & WISE J050434.26-734927.2 & Gal & 0.045 \\
XMMSL1 J050956.9-641741 & Y & 0.74$\pm$0.23 & 77.4886 & -64.2949 & GALEX J050957.2-641742 & QSO & - \\
XMMSL1 J051622.4+192701 & - & 1.27$\pm$0.31 & 79.0946 & 19.4531 & IRAS 05134+1923 & Gal & 0.021 \\
XMMSL1 J051949.5-454642 & Y & 0.59$\pm$0.20 & 79.9572 & -45.7789 & PICTOR A & Sy1 & 0.035 \\
XMMSL1 J053756.2-024516 & Y & 0.76$\pm$0.28 & 84.4846 & -2.7536 & WISE J053756.30-024513.1 & Gal & 0.58 \\
XMMSL1 J054357.2-553208 & Y & 0.81$\pm$0.22 & 85.9884 & -55.5354 & RBS 679 & BLLac & - \\
XMMSL1 J054641.6-641518 & Y & 1.20$\pm$0.35 & 86.6743 & -64.2561 & GALEX J054641.8-641521 & Sy1 & 0.323 \\
XMMSL1 J055040.6-321617 & Y & 2.67$\pm$0.44 & 87.6690 & -32.2713 & B0548-322 & QSO & 0.069 \\
XMMSL1 J055211.9-072723 & - & 5.37$\pm$0.89 & 88.0474 & -7.4562 & NGC 2110 & Sy2 & 0.0078 \\
XMMSL1 J060649.6-624547 & Y & 0.67$\pm$0.24 & 91.7070 & -62.7622 & GALEX J060649.6-624543 & UVX & - \\
XMMSL1 J061100.0-491036 & Y & 0.49$\pm$0.18 & 92.7507 & -49.1763 & GALEX J061100.1-491033 & QSO & - \\
XMMSL1 J062706.2-352916 & Y & 1.22$\pm$0.29 & 96.7780 & -35.4876 & PKS 0625-354 & LINER & 0.054 \\
XMMSL1 J063326.2-561416 & - & 0.59$\pm$0.20 & 98.3609 & -56.2392 & WISE J063326.61-561420.9 & Gal & 0.047 \\
XMMSL1 J063957.2+384840 & Y & 0.80$\pm$0.22 & 99.9894 & 38.8107 & B0636+3851 & QSO & - \\
XMMSL1 J071030.7+590807 & Y & 2.03$\pm$0.42 & 107.6253 & 59.1390 & B0706+5913 & BLLac & 0.125 \\
XMMSL1 J072927.7+243622 & Y & 0.60$\pm$0.22 & 112.3659 & 24.6066 & B2 0726+24 & Sy2 & 0.163 \\
XMMSL1 J072956.7-654338 & Y & 0.61$\pm$0.27 & 112.4878 & -65.7258 & 6dFGS g0729571-654333 & UVX & 0.08 \\
XMMSL1 J073956.1+280151 & - & 1.77$\pm$0.51 & 114.9844 & 28.0290 & GALEX J073956.2+280145 & UVX & 0.081 \\
XMMSL1 J074021.7-581544 & - & 0.40$\pm$0.16 & 115.0934 & -58.2634 & WISE J074022.41-581548.2 & Gal & - \\
XMMSL1 J074126.3+354716 & Y & 0.78$\pm$0.19 & 115.3599 & 35.7841 & GALEX J074126.3+354702 & QSO & - \\
XMMSL1 J075051.9+123101 & Y & 0.46$\pm$0.18 & 117.7169 & 12.5180 & B0748+126 & QSO & 0.889 \\
XMMSL1 J075118.9-665728 & - & 0.49$\pm$0.17 & 117.8269 & -66.9578 & WISE J075118.44-665727.9 & Gal & - \\
XMMSL1 J075244.5+455655 & Y & 0.68$\pm$0.20 & 118.1842 & 45.9493 & NPM1G+46.0092 & Sy1.9 & 0.051 \\
XMMSL1 J080157.9-494648 & Y & 1.21$\pm$0.40 & 120.4915 & -49.7785 & ESO 209-12 & Sy1 & 0.04 \\
XMMSL1 J080327.6+084156 & Y & 1.67$\pm$0.29 & 120.8641 & 8.6979 & FIRST J080327.4+084152 & Gal & 0.047 \\
XMMSL1 J081917.6-075626 & Y & 1.21$\pm$0.44 & 124.8233 & -7.9406 & RX J0819.2-0756 & BLLac & - \\
XMMSL1 J083014.8-094505 & Y & 0.86$\pm$0.21 & 127.5631 & -9.7488 & GALEX J083015.1-094456 & QSO & - \\

\hline                                   
\end{tabular}
\end{center}
\end{table*}

\setcounter{table}{0}

\begin{table*}
  \caption{Continued}
  \begin{center}
\begin{tabular}{llrrrllr}   
\hline
Slew ID & Soft & Flux & RA & Dec & ID & Type & Redshift \\
\hline

XMMSL1 J083015.3-672527 & Y & 0.67$\pm$0.17 & 127.5690 & -67.4248 & GALEX J083016.5-672528 & UVX & 0.035 \\
XMMSL1 J083950.7-121434 & Y & 0.76$\pm$0.27 & 129.9608 & -12.2429 & PKS 0837-12 & Sy1.2 & 0.197 \\
XMMSL1 J084314.1+535706 & Y & 0.61$\pm$0.24 & 130.8059 & 53.9553 & SBS 0839+541 & Sy1 & 0.218 \\
XMMSL1 J085036.8+044354 & Y & 0.88$\pm$0.32 & 132.6548 & 4.7326 & GALEX J085037.1+044356 & QSO & - \\
XMMSL1 J085454.2-242337 & Y & 1.14$\pm$0.34 & 133.7299 & -24.3937 & GALEX J085455.1-242337 & UVX & 0.091 \\
XMMSL1 J085841.0+104134 & - & 0.45$\pm$0.17 & 134.6740 & 10.6895 & IRAS 08559+1053 & Sy2 & 0.149 \\
XMMSL1 J090816.9+052012 & Y & 0.51$\pm$0.20 & 137.0715 & 5.3338 & SDSS J090816.78+052012.0 & QSO & 0.344 \\
XMMSL1 J090822.3-643749 & Y & 1.39$\pm$0.34 & 137.0937 & -64.6311 & WISE J090822.49-643751.7 & Gal & - \\
XMMSL1 J091300.4-210324 & Y & 2.24$\pm$0.51 & 138.2509 & -21.0559 & 6dFGS gJ091300.2-210321 & BLLac & 0.198 \\
XMMSL1 J092342.9+225435 & Y & 1.24$\pm$0.35 & 140.9292 & 22.9091 & CGCG 121-075 & Sy1.2 & 0.032 \\
XMMSL1 J092616.5-842131 & - & 0.71$\pm$0.23 & 141.5740 & -84.3594 & IRAS 09305-8408 & Sy2 & 0.062 \\
XMMSL1 J093909.6-211252 & Y & 0.50$\pm$0.18 & 144.7907 & -21.2141 & GALEX J093909.7-211250 & QSO & - \\
XMMSL1 J094745.3+072521 & - & 0.38$\pm$0.15 & 146.9381 & 7.4224 & 3C 227 & Sy1 & 0.087 \\
XMMSL1 J095303.0-765802 & Y & 0.70$\pm$0.25 & 148.2683 & -76.9673 & SUMSS J095303-765804 & Gal & - \\
XMMSL1 J095534.3+690350 & Y & 1.24$\pm$0.32 & 148.8883 & 69.0654 & M 81 & LINER & 1.4E-4 \\
XMMSL1 J095942.7-311255 & Y & 1.12$\pm$0.26 & 149.9277 & -31.2162 & 1RXS J095942.1-311300 & Sy1 & 0.037 \\
XMMSL1 J100207.0+030333 & Y & 0.64$\pm$0.16 & 150.5293 & 3.0577 & IC 588 & QSO & 0.023 \\
XMMSL1 J100955.8+261129 & - & 0.96$\pm$0.32 & 152.4848 & 26.1922 & GALEX J100956.1+261132 & UVX & 0.241 \\
XMMSL1 J102330.4+195157 & Y & 2.77$\pm$0.41 & 155.8774 & 19.8651 & NGC 3227 & Sy1.5 & 0.0039 \\
XMMSL1 J102356.0-433610 & Y & 1.45$\pm$0.27 & 155.9842 & -43.6004 & RXS J10239-4336 & BLLac & - \\
XMMSL1 J104833.4-390244 & Y & 1.02$\pm$0.31 & 162.1410 & -39.0439 & GALEX J104833.8-390237 & UVX & 0.044 \\
XMMSL1 J105639.8-312213 & Y & 0.41$\pm$0.17 & 164.1659 & -31.3700 & GALEX J105639.8-312211 & QSO & - \\
XMMSL1 J110427.8+381230 & Y & 5.97$\pm$0.73 & 166.1139 & 38.2089 & Mrk 421 & BLLac & 0.03 \\
XMMSL1 J112048.4+421212 & Y & 1.39$\pm$0.46 & 170.2003 & 42.2035 & BWE 1118+4228 & BLLac & 0.124 \\
XMMSL1 J112654.2+185007 & - & 0.35$\pm$0.15 & 171.7295 & 18.8326 & MCG +03-29-049 & RG & 0.019 \\
XMMSL1 J112715.8+190917 & Y & 1.32$\pm$0.27 & 171.8180 & 19.1556 & 1RXS J112716.6+190914 & Sy1 & 0.1 \\
XMMSL1 J112736.7+244918 & Y & 0.76$\pm$0.25 & 171.9038 & 24.8232 & GALEX J112736.9+244924 & QSO & 0.059 \\
XMMSL1 J112841.5+575017 & Y & 0.35$\pm$0.14 & 172.1709 & 57.8352 & MCG +10-17-004 & Sy2 & 0.051 \\
XMMSL1 J113104.0+685156 & Y & 0.92$\pm$0.49 & 172.7699 & 68.8647 & EXO 1128.1+6908 & Sy1 & 0.043 \\
XMMSL1 J113626.2+700926 & Y & 1.43$\pm$0.41 & 174.1101 & 70.1575 & B1136+704 & BLLac & 0.045 \\
XMMSL1 J120058.3+064828 & - & 0.93$\pm$0.29 & 180.2415 & 6.8064 & WISE J120057.95+064823.0 & Gal & 0.036 \\
XMMSL1 J121044.4+382010 & - & 0.50$\pm$0.19 & 182.6845 & 38.3362 & KUG 1208+386 & Sy1 & 0.023 \\
XMMSL1 J121158.5+224235 & Y & 0.67$\pm$0.25 & 182.9943 & 22.7092 & FIRST J121158.6+224232 & BLLac & 0.455 \\
XMMSL1 J121749.9+354454 & Y & 0.44$\pm$0.19 & 184.4575 & 35.7471 & GALEX J121749.7+354448 & UVX & 0.088 \\
XMMSL1 J121826.6+294846 & Y & 1.24$\pm$0.31 & 184.6105 & 29.8130 & Mrk 766 & Sy1 & 0.012 \\
XMMSL1 J122121.9+301040 & Y & 1.38$\pm$0.45 & 185.3415 & 30.1770 & B1218+304 & BLLac & 0.182 \\
XMMSL1 J122324.3+024052 & Y & 1.87$\pm$0.37 & 185.8506 & 2.6791 & Mrk 50 & Sy1.2 & 0.023 \\
XMMSL1 J122546.2+123930 & Y & 1.92$\pm$0.40 & 186.4449 & 12.6622 & NGC 4388 & Sy2 & 0.0084 \\
XMMSL1 J123135.7+704413 & Y & 0.50$\pm$0.18 & 187.9024 & 70.7374 & RX J1231.6+7044 & Sy1 & 0.208 \\
XMMSL1 J123203.4+200930 & Y & 1.24$\pm$0.33 & 188.0151 & 20.1582 & Mrk 771 & Sy1 & 0.063 \\
XMMSL1 J125212.3-132455 & Y & 1.44$\pm$0.42 & 193.0520 & -13.4147 & NGC 4748 & Sy1 & 0.014 \\
XMMSL1 J125341.2-393156 & Y & 0.74$\pm$0.19 & 193.4219 & -39.5332 & SHBL J125341.1-393159 & QSO & 0.19 \\
XMMSL1 J125456.5-265704 & Y & 0.76$\pm$0.19 & 193.7348 & -26.9505 & AM 1252-264 & Sy1 & 0.059 \\
XMMSL1 J125611.0-054720 & Y & 1.28$\pm$0.35 & 194.0466 & -5.7893 & 3C 279 & QSO & 0.536 \\
XMMSL1 J125616.0-114632 & Y & 0.81$\pm$0.27 & 194.0665 & -11.7770 & 6dFGS g1256160-114637 & Gal & 0.058 \\
XMMSL1 J125657.1-095017 & - & 1.22$\pm$0.46 & 194.2370 & -9.8374 & HE 1254-0934 & Sy1 & 0.139 \\
XMMSL1 J130258.9+162427 & Y & 1.32$\pm$0.35 & 195.7452 & 16.4077 & Mrk 783 & Sy1 & 0.067 \\
XMMSL1 J130323.0-134132 & Y & 0.45$\pm$0.19 & 195.8427 & -13.6925 & NPM1G-13.0407 & Sy1.2 & 0.046 \\
XMMSL1 J130359.4+033932 & - & 0.51$\pm$0.22 & 195.9978 & 3.6590 & FIRST J130359.5+033932 & Sy1 & 0.184 \\
XMMSL1 J130708.7+342417 & Y & 0.59$\pm$0.21 & 196.7866 & 34.4061 & Mrk 64 & Sy1.2 & 0.185 \\
XMMSL1 J131305.8-110745 & Y & 0.92$\pm$0.24 & 198.2741 & -11.1284 & II SZ 10 & Sy1.5 & 0.034 \\
XMMSL1 J131424.0+271919 & Y & 0.74$\pm$0.24 & 198.5981 & 27.3222 & CASG 991 & Sy1 & 0.133 \\
XMMSL1 J131516.5+442427 & - & 0.79$\pm$0.30 & 198.8222 & 44.4072 & UGC 8327 & QSO & 0.036 \\
XMMSL1 J133239.1-102853 & Y & 0.64$\pm$0.22 & 203.1630 & -10.4812 & MCG -02-35-001 & Sy1 & 0.022 \\
XMMSL1 J133553.1-341745 & Y & 2.52$\pm$0.49 & 203.9741 & -34.2956 & MCG -06-30-015 & Sy1.5 & 0.0070 \\
XMMSL1 J133718.8+242258 & Y & 0.50$\pm$0.17 & 204.3280 & 24.3843 & B1334+2438 & Sy1 & 0.107 \\
XMMSL1 J133844.1+242945 & - & 1.03$\pm$0.41 & 204.6814 & 24.5009 & WISE J133843.54+243003.0 & QSO & 0.055 \\
XMMSL1 J134105.2+395946 & Y & 0.88$\pm$0.20 & 205.2713 & 39.9960 & 7C 1338+4015 & BLLac & 0.163 \\
XMMSL1 J134356.6+253849 & Y & 0.60$\pm$0.27 & 205.9865 & 25.6466 & B1341+258 & Sy1 & 0.087 \\
XMMSL1 J134442.0-451002 & - & 0.79$\pm$0.29 & 206.1741 & -45.1687 & PGC 529701 & Gal & 0.145 \\
XMMSL1 J134519.4+414243 & Y & 0.97$\pm$0.28 & 206.3299 & 41.7124 & NGC 5290 & QSO & 0.0085 \\
XMMSL1 J134915.2+220029 & Y & 0.49$\pm$0.19 & 207.3134 & 22.0091 & IRAS F13469+2215 & Sy1 & 0.062 \\
XMMSL1 J135304.2+691826 & Y & 1.66$\pm$0.52 & 208.2644 & 69.3082 & B1351+695 & Sy1 & 0.03 \\
XMMSL1 J140743.6-430516 & Y & 1.27$\pm$0.31 & 211.9323 & -43.0858 & WISE J140743.75-430508.8 & Gal & - \\
XMMSL1 J140806.7-302348 & Y & 0.49$\pm$0.18 & 212.0283 & -30.3983 & EQ 1405-301 & Sy1 & 0.023 \\

\hline                                   
\end{tabular}
\end{center}
\end{table*}

\setcounter{table}{0}

\begin{table*}
  \caption{Continued}
  \begin{center}
\begin{tabular}{llrrrllr}   
\hline
Slew ID & Soft & Flux & RA & Dec & ID & Type & Redshift \\ 
\hline

XMMSL1 J141117.1-340323 & Y & 0.69$\pm$0.29 & 212.8238 & -34.0560 & 6dFGS g1411177-340321 & UVX & 0.037 \\
XMMSL1 J141922.4-263841 & Y & 3.46$\pm$0.56 & 214.8434 & -26.6447 & ESO 511-030 & Sy1 & 0.022 \\
XMMSL1 J142129.5+474724 & Y & 0.72$\pm$0.25 & 215.3740 & 47.7902 & SBS 1419+480 & Sy1.5 & 0.072 \\
XMMSL1 J142149.3-380901 & Y & 0.94$\pm$0.29 & 215.4628 & -38.1483 & WISE J142151.07-380853.7 & Gal & - \\
XMMSL1 J142239.3+580156 & Y & 1.17$\pm$0.31 & 215.6620 & 58.0321 & QSO B1422+580 & BLLac & 0.634 \\
XMMSL1 J143342.1-730442 & Y & 0.76$\pm$0.20 & 218.4286 & -73.0772 & GALEX J143343.0-730437 & QSO & - \\
XMMSL1 J144923.5-063835 & Y & 0.86$\pm$0.30 & 222.3472 & -6.6471 & PGC 1030451 & Gal & 0.085 \\
XMMSL1 J144932.5+274624 & Y & 0.89$\pm$0.27 & 222.3862 & 27.7728 & RBS 1434 & BLLac & 0.225 \\
XMMSL1 J145108.6+270928 & Y & 0.94$\pm$0.38 & 222.7866 & 27.1575 & QSO B1448+273 & Sy1 & 0.065 \\
XMMSL1 J145307.4+255432 & Y & 2.10$\pm$0.38 & 223.2830 & 25.9092 & GALEX J145307.9+255433 & QSO & 0.048 \\
XMMSL1 J145907.4+714011 & - & 1.01$\pm$0.29 & 224.7817 & 71.6722 & 3C 309.1 & Sy1 & 0.905 \\
XMMSL1 J150401.2+102618 & Y & 2.45$\pm$0.75 & 226.0050 & 10.4378 & Mrk 841 & Sy1.5 & 0.036 \\
XMMSL1 J150745.4+512716 & - & 0.66$\pm$0.28 & 226.9375 & 51.4529 & Mrk 845 & Sy1 & 0.046 \\
XMMSL1 J151618.8-152345 & Y & 0.35$\pm$0.16 & 229.0780 & -15.3956 & 1RXS J151618.7-152347 & BLLac & - \\
XMMSL1 J153500.6+532035 & Y & 0.77$\pm$0.22 & 233.7534 & 53.3437 & B1533+535 & BLLac & 0.89 \\
XMMSL1 J153552.4+575415 & Y & 1.39$\pm$0.29 & 233.9684 & 57.9027 & Mrk 290 & Sy1 & 0.029 \\
XMMSL1 J154016.3+815507 & Y & 1.93$\pm$0.54 & 235.0663 & 81.9182 & 1ES 1544+820 & BLLac & 0.271 \\
XMMSL1 J154631.4-282217 & Y & 0.55$\pm$0.21 & 236.6306 & -28.3695 & GALEX J154631.3-282210 & RG & 0.121 \\
XMMSL1 J154815.8+694942 & Y & 0.56$\pm$0.24 & 237.0697 & 69.8265 & RX J1548.3+6949 & QSO & 0.375 \\
XMMSL1 J154824.8-134526 & Y & 1.27$\pm$0.26 & 237.1040 & -13.7576 & NGC 5995 & Sy1.9 & 0.025 \\
XMMSL1 J155542.9+111126 & Y & 2.38$\pm$0.51 & 238.9294 & 11.1901 & PG 1553+11 & BLLac & 0.36 \\
XMMSL1 J161124.7+585102 & Y & 0.52$\pm$0.15 & 242.8529 & 58.8504 & SBS 1610+589 & Sy1.5 & 0.032 \\
XMMSL1 J161126.8+724027 & - & 0.51$\pm$0.23 & 242.8624 & 72.6745 & WISE J161126.98+724028.1 & Gal & - \\
XMMSL1 J161658.4+643845 & Y & 0.45$\pm$0.19 & 244.2428 & 64.6450 & HS 1616+6445 & Sy1 & 0.171 \\
XMMSL1 J161752.8-771723 & Y & 0.56$\pm$0.23 & 244.4628 & -77.2929 & QSO B1610-771 & QSO & 1.71 \\
XMMSL1 J162532.1+852949 & Y & 0.91$\pm$0.29 & 246.3583 & 85.4949 & VII Zw 653 & Sy1.2 & 0.063 \\
XMMSL1 J162948.1+672245 & Y & 0.71$\pm$0.32 & 247.4516 & 67.3784 & Mrk 0885 & Sy1.5 & 0.052 \\
XMMSL1 J164734.8+495013 & Y & 0.53$\pm$0.19 & 251.8944 & 49.8387 & RXS J16475+4950 & Sy1 & 0.047 \\
XMMSL1 J165218.8+555415 & - & 0.60$\pm$0.17 & 253.0787 & 55.9056 & UGC 10593 & Sy2 & 0.029 \\
XMMSL1 J170025.5-724045 & Y & 0.57$\pm$0.21 & 255.1059 & -72.6791 & GALEX J170025.1-724044 & UVX & 0.104 \\
XMMSL1 J170330.6+454041 & Y & 0.49$\pm$0.16 & 255.8766 & 45.6798 & B3 1702+457 & Sy1 & 0.061 \\
XMMSL1 J170500.6-013227 & Y & 0.93$\pm$0.34 & 256.2516 & -1.5413 & UGC 10683 & Sy1 & 0.03 \\
XMMSL1 J172504.4+115213 & Y & 1.39$\pm$0.35 & 261.2681 & 11.8710 & QSO B1722+119 & BLLac & 0.018 \\
XMMSL1 J172700.3+181422 & Y & 0.55$\pm$0.17 & 261.7535 & 18.2390 & GALEX J172700.7+181420 & QSO & - \\
XMMSL1 J173336.0+363130 & - & 0.91$\pm$0.33 & 263.4037 & 36.5255 & WISE J173336.88+363131.8 & Gal & - \\
XMMSL1 J174037.2+521139 & Y & 0.50$\pm$0.19 & 265.1541 & 52.1954 & QSO B1739+522 & QSO & 1.375 \\
XMMSL1 J174209.3+510107 & Y & 0.66$\pm$0.28 & 265.5385 & 51.0181 & GALEX J174209.2+510105 & UVX & 0.063 \\
XMMSL1 J174357.7+193511 & Y & 1.93$\pm$0.37 & 265.9910 & 19.5859 & B1741+196 & BLLac & 0.083 \\
XMMSL1 J174415.0+325930 & Y & 1.13$\pm$0.32 & 266.0603 & 32.9915 & 87 GB174224.8+3300 & Gal & 0.076 \\
XMMSL1 J174508.2+015448 & - & 0.45$\pm$0.21 & 266.2826 & 1.9118 & WISE J174507.82+015442.5 & Gal & - \\
XMMSL1 J174538.1+290820 & Y & 1.10$\pm$0.31 & 266.4095 & 29.1395 & GALEX J174538.2+290822 & Sy1 & 0.111 \\
XMMSL1 J175104.6+744742 & Y & 0.48$\pm$0.20 & 267.7701 & 74.7956 & PGC 60979 & Gal & 0.025 \\
XMMSL1 J180058.3+541035 & Y & 0.56$\pm$0.19 & 270.2436 & 54.1745 & GALEX J180058.5+541027 & UVX & - \\
XMMSL1 J180951.8-655615 & - & 0.70$\pm$0.21 & 272.4679 & -65.9371 & PMN J1809-6556 & Gal & 0.18 \\
XMMSL1 J182021.1+362339 & Y & 0.42$\pm$0.16 & 275.0875 & 36.3953 & GALEX J182021.0+362344 & QSO & - \\
XMMSL1 J182609.9+545005 & Y & 0.94$\pm$0.29 & 276.5457 & 54.8349 & WISE J182610.95+545005.7 & Gal & - \\
XMMSL1 J183503.4+324143 & Y & 4.49$\pm$0.81 & 278.7641 & 32.6964 & 4C 32.55 & Sy1 & 0.057 \\
XMMSL1 J183658.3-592405 & Y & 2.09$\pm$0.50 & 279.2426 & -59.4023 & FRL 49 & Sy2 & 0.02 \\
XMMSL1 J185649.5-544233 & Y & 1.92$\pm$0.55 & 284.2057 & -54.7083 & 6dFGS g1856494-544230 & UVX & 0.056 \\
XMMSL1 J190939.2-622852 & Y & 1.89$\pm$0.42 & 287.4120 & -62.4820 & ESO 104-41 & UVX & 0.082 \\
XMMSL1 J191028.2+495606 & Y & 0.44$\pm$0.16 & 287.6186 & 49.9348 & WISE J191028.46+495605.4 & Gal & 0.146 \\
XMMSL1 J191921.7+502646 & Y & 0.53$\pm$0.22 & 289.8421 & 50.4463 & PGC 2373270 & UVX & 0.066 \\
XMMSL1 J192404.3+552933 & Y & 0.39$\pm$0.17 & 291.0198 & 55.4939 & GALEX J192404.7+552937 & QSO & - \\
XMMSL1 J194240.3-101919 & Y & 2.58$\pm$0.65 & 295.6691 & -10.3236 & NGC 6814 & Sy1 & 0.0052 \\
XMMSL1 J201117.1+600431 & Y & 1.14$\pm$0.30 & 302.8201 & 60.0744 & GALEX J201116.8+600428 & QSO & - \\
XMMSL1 J202931.2-614858 & - & 0.62$\pm$0.20 & 307.3801 & -61.8191 & PGC 352109 & Gal & 0.19 \\
XMMSL1 J204142.3+185258 & Y & 1.25$\pm$0.40 & 310.4283 & 18.8842 & GALEX J204142.8+185303 & QSO & - \\
XMMSL1 J204206.1+242653 & Y & 0.78$\pm$0.27 & 310.5252 & 24.4479 & 1RXS J204206.3+242655 & BLLac & 0.102 \\
XMMSL1 J204236.3+240615 & - & 0.71$\pm$0.33 & 310.6491 & 24.1026 & WISE J204235.78+240609.1 & Gal & - \\
XMMSL1 J205129.5-154510 & - & 0.49$\pm$0.16 & 312.8729 & -15.7533 & GALEX J205129.3-154511 & QSO & - \\
XMMSL1 J205558.4-152927 & - & 0.56$\pm$0.19 & 313.9914 & -15.4914 & PGC 3093850 & Gal & 0.08 \\
XMMSL1 J205617.1-471450 & Y & 0.76$\pm$0.23 & 314.0681 & -47.2465 & B2052-474 & QSO & 1.489 \\
XMMSL1 J211354.1+820450 & Y & 2.31$\pm$0.46 & 318.5048 & 82.0801 & S5 2116+81 & Sy1 & 0.086 \\
XMMSL1 J211928.2+333305 & Y & 1.61$\pm$0.45 & 319.8714 & 33.5492 & WISE J211929.13+333257.1 & Gal & 0.051 \\
XMMSL1 J212434.9-113122 & - & 0.48$\pm$0.16 & 321.1449 & -11.5232 & 1RXS J212434.7-113125 & Gal & 0.42 \\

\hline                                   
\end{tabular}
\end{center}
\end{table*}

\setcounter{table}{0}

\begin{table*}
  \caption{Continued}
  \begin{center}
\begin{tabular}{llrrrllr}   
\hline
Slew ID & Soft & Flux & RA & Dec & ID & Type & Redshift \\ 
\hline

XMMSL1 J212940.5+000518 & Y & 0.82$\pm$0.26 & 322.4166 & 0.0894 & WISE J212939.98+000521.9 & Gal & 0.234 \\
XMMSL1 J213227.6+100821 & Y & 0.87$\pm$0.31 & 323.1159 & 10.1387 & UGC 11763 & Sy1 & 0.063 \\
XMMSL1 J214958.0-185923 & Y & 0.81$\pm$0.23 & 327.4919 & -18.9899 & RXS J21499-1859 & Sy1 & 0.158 \\
XMMSL1 J215155.4-302755 & Y & 0.75$\pm$0.24 & 327.9813 & -30.4649 & B2149-306 & QSO & 2.345 \\
XMMSL1 J215705.4+063222 & - & 0.57$\pm$0.20 & 329.2729 & 6.5381 & PGC 3091877 & Gal & 0.115 \\
XMMSL1 J220907.9-274843 & Y & 1.11$\pm$0.32 & 332.2820 & -27.8095 & NGC 7214 & Sy1.2 & 0.023 \\
XMMSL1 J220915.8-470954 & Y & 2.09$\pm$0.36 & 332.3175 & -47.1667 & NGC 7213 & Sy1 & 0.0060 \\
XMMSL1 J222344.1+115010 & Y & 0.69$\pm$0.21 & 335.9376 & 11.8359 & MCG +02-57-002 & Sy1.5 & 0.029 \\
XMMSL1 J222547.2-045701 & - & 0.81$\pm$0.32 & 336.4469 & -4.9503 & 4C -05.92 & BLLac & 1.404 \\
XMMSL1 J222940.2-083253 & Y & 0.83$\pm$0.32 & 337.4170 & -8.5484 & PKS 2227-088 & QSO & 1.559 \\
XMMSL1 J223248.4-202222 & Y & 0.63$\pm$0.18 & 338.2033 & -20.3739 & GALEX J223248.8-202225 & QSO & - \\
XMMSL1 J223301.0+133557 & Y & 0.85$\pm$0.32 & 338.2546 & 13.6006 & RXS J2233.0+1335 & BLLac & 0.213 \\
XMMSL1 J223546.3-260301 & Y & 1.47$\pm$0.31 & 338.9425 & -26.0504 & NGC 7314 & Sy1.9 & 0.0047 \\
XMMSL1 J224641.8-520637 & Y & 0.85$\pm$0.24 & 341.6754 & -52.1112 & RBS 1895 & BLLac & 0.194 \\
XMMSL1 J225100.7+781447 & Y & 0.62$\pm$0.22 & 342.7545 & 78.2445 & WISE J225101.09+781440.1 & Gal & - \\
XMMSL1 J225146.8-320613 & Y & 1.12$\pm$0.22 & 342.9481 & -32.1036 & RBS 1906 & BLLac & 0.246 \\
XMMSL1 J225505.8-020628 & - & 0.99$\pm$0.39 & 343.7771 & -2.1080 & WISE J225506.50-020628.7 & Gal & - \\
XMMSL1 J225932.6+245502 & Y & 0.62$\pm$0.21 & 344.8872 & 24.9184 & KAZ 320 & Sy1 & 0.034 \\
XMMSL1 J230047.1-125513 & Y & 0.56$\pm$0.23 & 345.1992 & -12.9185 & NGC 7450 & Sy1 & 0.01 \\
XMMSL1 J230302.8-184128 & Y & 0.73$\pm$0.22 & 345.7624 & -18.6905 & PKS 2300-18 & Sy1 & 0.128 \\
XMMSL1 J230402.2+223726 & Y & 0.39$\pm$0.16 & 346.0109 & 22.6243 & Mrk 315 & Sy1 & 0.039 \\
XMMSL1 J232523.8-382651 & Y & 1.57$\pm$0.51 & 351.3508 & -38.4470 & IRAS 23226-3843 & Sy1 & 0.036 \\
XMMSL1 J234333.9+343955 & Y & 0.82$\pm$0.30 & 355.8899 & 34.6641 & SHBL J234333.8+344004 & QSO & 0.366 \\
XMMSL1 J235727.7-302738 & Y & 1.42$\pm$0.60 & 359.3668 & -30.4612 & DUGRS 471-011 & Sy1 & 0.03 \\

\hline                                   
\end{tabular}
\end{center}
\end{table*}

\end{document}